\documentclass[manuscript,screen,nonacm]{acmart}

\usepackage{graphicx}
\usepackage{balance}
\usepackage{booktabs}
\usepackage{url}
\usepackage{wrapfig}

\usepackage[subtle]{savetrees}

\usepackage[font=bf,labelfont=bf]{caption}
\usepackage{subcaption}

\usepackage[linesnumbered,ruled,vlined,algo2e]{algorithm2e}

\SetCommentSty{mycommfont}

\usepackage{xspace}

\newcommand{\beforeCaptionSpacing}{\vspace{-0.2cm}}
\newcommand{\afterCaptionSpacing}{\vspace{-0.2cm}}

\usepackage{xcolor}
\usepackage{colortbl}

\SetKwInOut{Input}{Input}
\SetKwInOut{Output}{Output}
\SetKw{return}{return}
\SetKwFunction{memcpy}{memcpy}
\SetKwFunction{sizeof}{sizeof}
\SetKwFunction{range}{range}

\SetKwFunction{buildindex}{BuildLearnedIndex}
\SetKwFunction{sort}{Sort}
\SetKwFunction{partition}{Partition}

\SetKwFunction{localsearch}{LocalSearch}
\SetKwFunction{searchpoint}{SearchPoint}
\SetKwFunction{indexlookup}{IndexLookup}
\SetKwFunction{estimatefrom}{EstimateFrom}
\SetKwFunction{estimateto}{EstimateTo}

\SetKwFunction{getmbr}{GetMBR}
\SetKwFunction{rangequery}{RangeQuery}
\SetKwFunction{withindistance}{WithinDistance}
\SetKwFunction{contains}{Contains}

\definecolor{lightgray}{gray}{0.90}
\newcommand\greybox[1]{%
  \vskip0.1\baselineskip%
  \par\noindent\colorbox{lightgray}{%
    \begin{minipage}{\columnwidth}#1\end{minipage}%
  }%
}

\usepackage{subfiles}

\definecolor{hmcolor1}{RGB}{254,237,222}
\definecolor{hmcolor2}{RGB}{253,208,162} 
\definecolor{hmcolor3}{RGB}{253,174,107} 
\definecolor{hmcolor4}{RGB}{253,141,60}
\definecolor{hmcolor5}{RGB}{230,85,13}
\definecolor{hmcolor6}{RGB}{166,54,3}

\title[Enhancing In-Memory Spatial Indexing with Learned Search]{Enhancing In-Memory Spatial Indexing with Learned Search}

\begin{document}

\author{Varun Pandey}
\authornote{These authors contributed equally to this work.}
\orcid{}
\affiliation{
  \institution{TU Berlin}
  \city{Berlin}
  \country{Germany}
}
\email{varun.pandey@tu-berlin.de}

\author{Alexander van Renen}
\authornotemark[1]
\authornote{Work done prior to joining Amazon.}
\affiliation{
  \institution{Amazon Web Services}
  \city{Munich}
  \country{Germany}
}
\email{vanralex@amazon.com}

\author{Eleni {Tzirita Zacharatou}}

\affiliation{
  \institution{IT University of Copenhagen}
  \city{Copenhagen}
  \country{Denmark}
}
\email{elza@itu.dk}

\author{Andreas Kipf}
\authornotemark[2]
\affiliation{
  \institution{Amazon Web Services}
  \city{Munich}
  \country{Germany}
}
\email{kipf@amazon.com}

\author{Ibrahim Sabek}
\affiliation{
  \institution{University of Southern California}
  \city{Los Angeles}
  \country{USA}
}
\email{sabek@usc.edu}

\author{Jialin Ding}
\authornotemark[2]
\affiliation{
  \institution{Amazon Web Services}
  \city{Munich}
  \country{Germany}
}
\email{jialind@amazon.com}

\author{Volker Markl}
\affiliation{
  \institution{TU Berlin and}
  \institution{DFKI GmbH}
  \city{Berlin}
  \country{Germany}
}
\email{volker.markl@tu-berlin.de}

\author{Alfons Kemper}
\affiliation{
  \institution{TU Munich}
  \city{Munich}
  \country{Germany}
}
\email{kemper@in.tum.de}

\renewcommand{\shortauthors}{Pandey, van Renen, Tzirita Zacharatou, et al.}

\begin{abstract}Spatial data is ubiquitous. Massive amounts of data are generated every day from a plethora of sources such as billions of GPS-enabled devices (e.g., cell phones, cars, and sensors), consumer-based applications (e.g., Uber and Strava), and social media platforms (e.g., location-tagged posts on Facebook, Twitter, and Instagram). 
This exponential growth in spatial data has led the research community to build systems and applications for efficient spatial data processing.

In this study, we apply a recently developed machine-learned search technique for single-dimensional sorted data to spatial indexing.
Specifically, we partition spatial data using six traditional spatial partitioning techniques and employ machine-learned search within each partition to support point, range, distance, and spatial join queries. Adhering to the latest research trends, we tune the partitioning techniques to be instance-optimized. By tuning each partitioning technique for optimal performance, we demonstrate that:
(i) grid-based index structures outperform tree-based index structures (from 1.23$\times$ to 2.47$\times$),
(ii) learning-enhanced variants of commonly used spatial index structures outperform their original counterparts (from 1.44$\times$ to 53.34$\times$ faster),
(iii) machine-learned search within a partition is faster than binary search by 11.79\% - 39.51\% when filtering on one dimension, 
(iv) the benefit of machine-learned search diminishes in the presence of other compute-intensive operations (e.g. \emph{scan} costs in higher selectivity queries, \emph{Haversine} distance computation, and \emph{point-in-polygon} tests), and
(v) index lookup is the bottleneck for tree-based structures, which could potentially be reduced by linearizing the indexed partitions. 

\end{abstract}

\keywords{spatial data, indexing, machine-learning, spatial queries, geospatial}

\maketitle

\newpage
\section{Introduction}
With the increase in the amount of spatial data available today, the database community has devoted substantial attention to spatial data management. 
For example, the NYC Taxi Rides open dataset~\cite{nyctaxidata} consists of pick-up and drop-off locations of more than 2.7 billion rides taken in the city since 2009. This represents more than 650,000 taxi rides per day in one of the most densely populated cities in the world but is only a fraction of the location data that is captured by many applications today. Uber, a popular ride-hailing service, operates on a global scale and completed 10 billion rides in 2018~\cite{uber_ten_billion}.
The unprecedented generation rate of location data has prompted a considerable amount of research efforts focused on scale-out systems~\cite{hadoopgis, distributed_data_store, spatialhadoop, sphinx, stark, locationspark, srx, hadoop_storage, simba, spatialspark, geospark, beast}, databases~\cite{memsql, databases_spatial_comparison, mongodb, oracle, hyperspace}, improving spatial query processing~\cite{hadoop_distance_joins, edbtjoin, geojoin, approxjoin, clipped_boxes, paralleljoin, main_memory_joins, two_level_spatial_index, bundled_queries, main_memory_spatial_temporal, geoblocks, bounded_approx, routing_1, incremental_spatial_partitioning, spatial_parquet_1, spatial_parquet_2, april}, or leveraging modern hardware and compiling techniques~\cite{gpufriendly_1, gpu_friendly_2, spatial_compiler_vision, spatial_compiler_2, spatial_compiler, gpurasterization, spade, spade_extended}, to handle the increasing demands of applications today.

Recently, Kraska et al.~\cite{rmi} proposed using learned models instead of traditional database indexes to predict the location of a key in a sorted dataset and demonstrated that they are typically faster than binary searches. Kester et al.~\cite{scanorprobe} showed that index scans are more efficient than optimized sequential scans in main-memory analytical engines for queries that select a small portion of the data. In this paper, we build on top of these recent research results and thoroughly investigate the impact of applying ideas from learned index structures (e.g., Flood~\cite{nathan2020flood}) on classical multidimensional indexes. 

\begin{wrapfigure}{r}{0.55\textwidth}
  \begin{center}
    \includegraphics[width=0.53\textwidth]{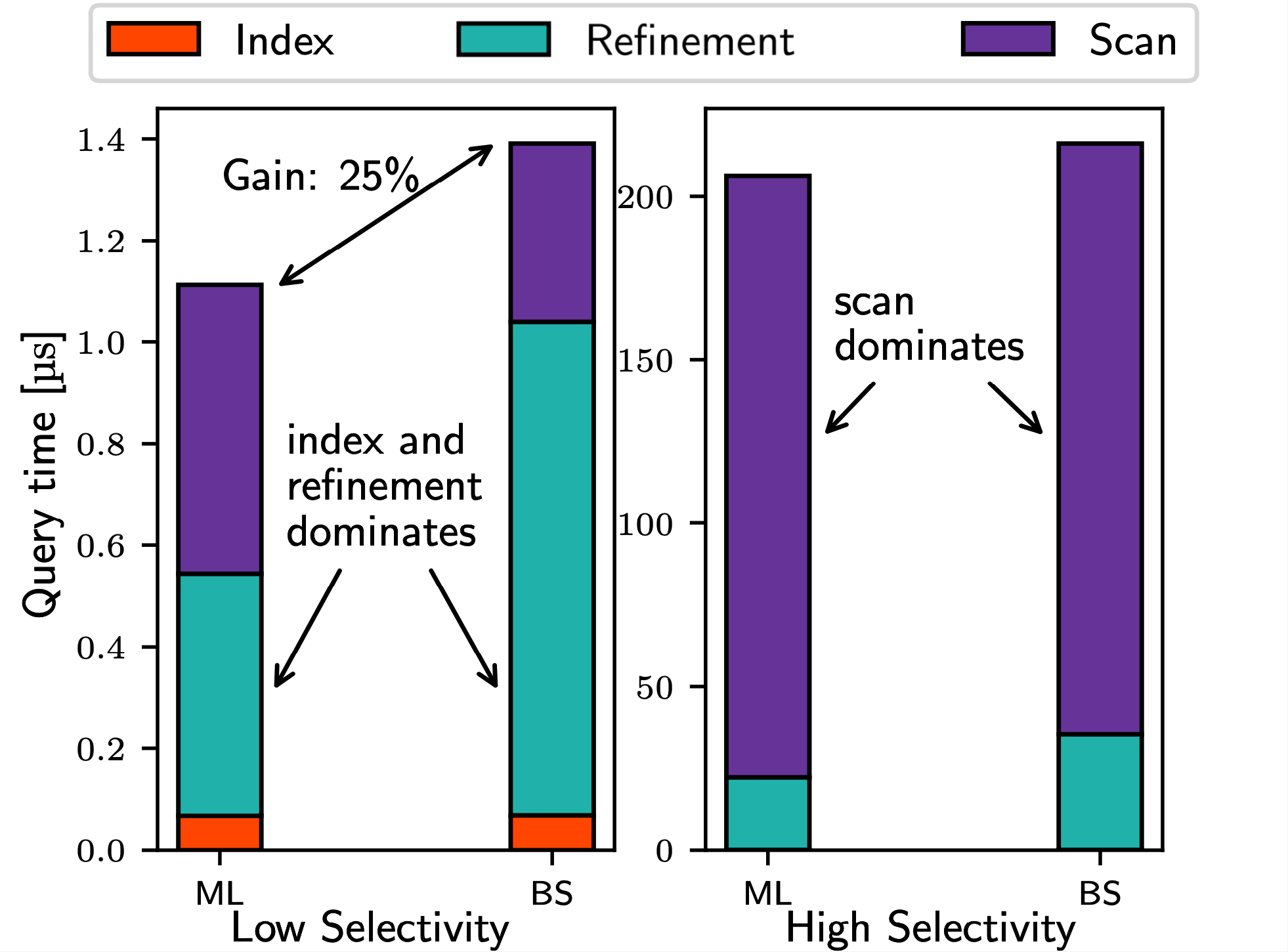}
  \end{center}
  \caption{Machine Learning vs. Binary Search (Spatial Range Query). For low selectivity (0.00001\%), the index and  boundary refinement phases dominate. For high selectivity (0.1\%), the scan phase dominates. Parameters are tuned to favor Binary Search.}
    \label{fig:mlvsbs}
\end{wrapfigure}

Specifically, we focus on six core spatial indexing techniques, namely linearization using Hilbert space-filling curve, fixed-grid~\cite{fixedgrid}, adaptive-grid~\cite{gridfile}, Kd-tree~\cite{kdtree}, Quadtree~\cite{quadtree} and STRtree~\cite{str_packing}. 
Query processing using these indexing techniques typically consists of three phases: index lookup, boundary refinement, and scanning. The index lookup phase identifies the intersecting partitions, the boundary refinement phase locates the lower bound of the query on the sorted dimension within the partition, and the scan phase scans the partition to find the qualifying points.
Section~\ref{sec:query_processing} provides more details on these phases.
In this paper, \emph{we propose using learned models, such as RadixSpline~\cite{radixspline}, to replace the traditional search techniques (e.g., binary search) used in the boundary refinement phase}.

Interestingly, we discovered that using a learned model as the search technique for boundary refinement can significantly improve query runtime, particularly for low-selectivity~\footnote{We adopt the definition of “selectivity” used by Pat Selinger et al.~\cite{access_path_selection_pat_sellinger}. Therefore, low selectivity indicates that the result set of a query has few qualifying tuples, while high selectivity indicates the opposite.} range queries (similar to the observation from Kester et al.~\cite{scanorprobe}). 
This applies to various queries, but the benefit decreases when other dominant costs such as scans, Haversine distance computations, and point-in-polygon tests are present. 
Figure~\ref{fig:mlvsbs} shows the average running time of a range query using adaptive-grid on a Tweets dataset, which consists of 83 million records (Section~\ref{sec:datasets} provides more details about the dataset), with and without learning. As shown in the figure, for a low-selectivity query (which selects 0.00001\% of the data, i.e., 8 records), the index and boundary refinement times dominate the lookup.
In contrast, for a high-selectivity query (which selects 0.1\% of the data, i.e., 83 thousand records) the scan time dominates. 
Additionally, our study found that one-dimensional grid partitioning techniques (e.g., fixed-grid) benefit more from the use of learned models than two-dimensional techniques (e.g., Quadtree).

We have also discovered that, contrary to conventional wisdom, grid-based indexes, which filter on one dimension and index on the other, are \emph{consistently} faster than tree-based indexes. This is because grid-based indexes typically have very large partitions for optimal performance, allowing for fast searches based on learned models within each partition. This advantage might not extend to disk-based index structures due to the confinement of partition size by page dimensions. We also note that another advantage of grid-based indexes is that they are the \emph{simplest} to implement.

In this paper, we extend the work presented in our previous publication~\cite{aidblearned}. 
In that previous work, we showed preliminary results \emph{only for range queries} using three datasets and \emph{five} learned indexes. 
In this study, we expand on our previous research by adding \emph{three more query types}, \emph{one more learned index}, and \emph{two competitive methods} commonly used in various applications and systems. Specifically, we extend the previous work as follows:
\begin{itemize}
  \item \textbf{Queries}: We implement \emph{three additional query types} (i.e., point, distance, and spatial join) in this work, bringing the total number of evaluated queries to four.
  \item \textbf{Index Structures}: We also \emph{add a learning-enhanced index based on linearization using the Hilbert curve} to the five previously evaluated learning-enhanced indexes (i.e., fixed-grid, adaptive-grid, K-d tree, Quadtree, and STRtree), bringing the total number of evaluated learning-enhanced indexes to six.
  \item \textbf{Competitors}: We further extend our preliminary study by including two index structures that are widely used in hundreds of applications and systems~\cite{spatiallibs, fullspatiallibs}. Concretely, we implement \emph{all four} queries using the JTS STRtree from the JTS library and the S2PointIndex from the Google S2 library.
\end{itemize}

We summarize the main findings of our experimental study as follows:
\begin{itemize}
  \item \textbf{Grid-based vs Tree-based}: Grid-based index structures, when tuned for optimal performance, outperform tree-based index structures due to (1) fewer random accesses, and (2) allowing for fast search using learned models over large partitions. They are up to 2.47$\times$ \emph{faster} compared to the closest tree-based competitor, exhibit \emph{robust} performance across various queries, and are also the \emph{simplest} to implement.
  \item \textbf{Effect of Learned Search}: We show that using learned models to search within partitions is 11.79\% to 39.51\% faster than binary search.
  \item \textbf{Compute-Intensive Operations}: The combined effect of grid-based indexes and learned models diminishes in the presence of compute-intensive operations, such as Haversine distance computations and point-in-polygon tests.
  \item \textbf{Performance Compared to Widely Used Indexes}: When compared to two commonly used index structures in various systems and applications, machine-learned indexes demonstrate superior performance, with speedups ranging from 1.44$\times$ to 53.34$\times$.
  
\end{itemize}

Our study sheds light on the effectiveness of different spatial index structures when used in conjunction with learned models. By evaluating multiple query types and index structures, our aim is to guide researchers and practitioners in choosing the best approach for their specific needs. Furthermore, our findings contribute to the design of improved spatial indexing methods using learned models.

\noindent \textbf{Outline.} The remainder of this paper is structured as follows. Section~\ref{sec:approach} presents the spatial indexing techniques that we implemented in our work, their learned variants, as well as the implementation of the different query types.
Then, Section~\ref{sec:evaluation} presents our experimental study. Finally, we discuss the related work in Section~\ref{sec:relatedwork} before concluding in Section~\ref{sec:conclusion}.

\section{Approach}
\label{sec:approach}

\begin{figure*}[htbp!]
	\centering
	\begin{subfigure}{0.3\textwidth}
		\centering
		\scalebox{.25}{\includegraphics{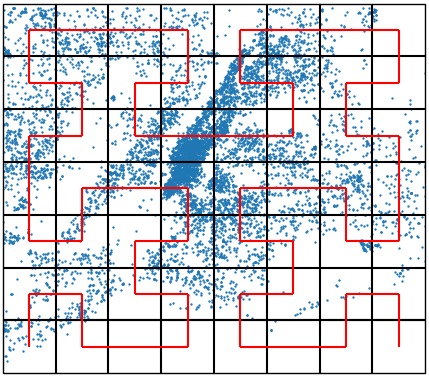}}
		\caption{Hilbert Space Filling Curve}
	\end{subfigure}
	\hspace{2mm}
	\begin{subfigure}{0.3\textwidth}
	    \centering
	    \scalebox{.24}{\includegraphics{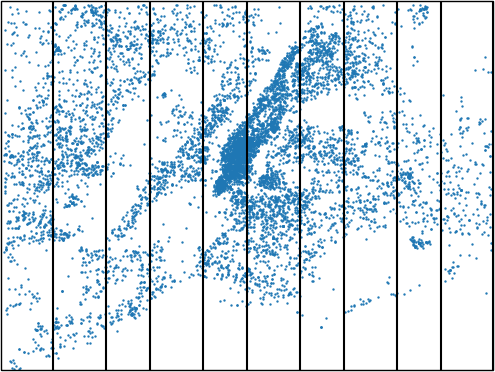}}
	    \caption{Fixed-grid}
	    \end{subfigure}
	\hspace{2mm}
	\begin{subfigure}{0.3\textwidth}
		\centering
		\scalebox{.24}{\includegraphics{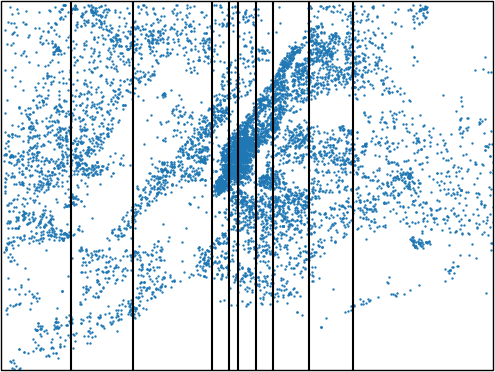}}
		\caption{Adaptive-grid}
	\end{subfigure}
	\hspace{2mm}
	\begin{subfigure}{0.3\textwidth}
		\centering
		\scalebox{.24}{\includegraphics{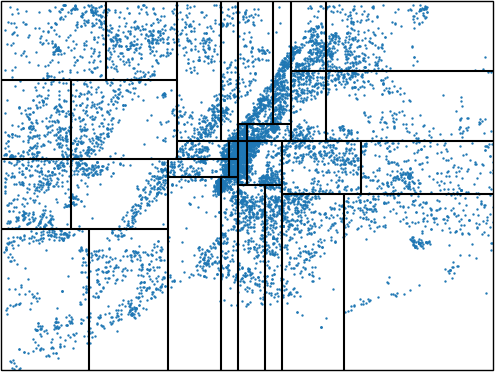}}
		\caption{K-d tree}
	\end{subfigure}
	\hspace{2mm}
	\begin{subfigure}{0.3\textwidth}
		\centering
		\scalebox{.235}{\includegraphics{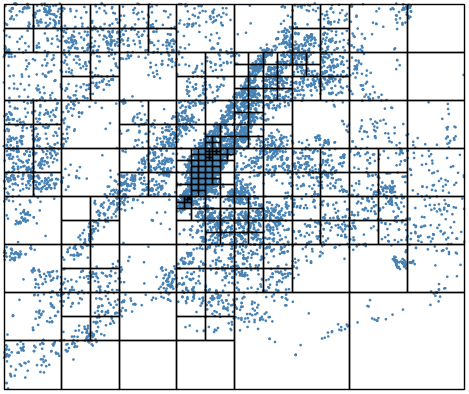}}
		\caption{Quadtree}
	\end{subfigure}
	\hspace{2mm}
	\begin{subfigure}{0.3\textwidth}
		\centering
		\scalebox{.24}{\includegraphics{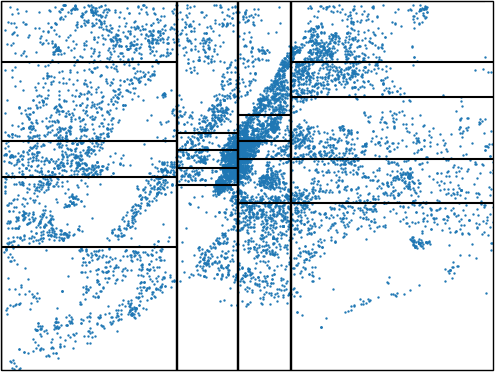}}
		\caption{STRtree}
	\end{subfigure}
	\beforeCaptionSpacing{}
	\caption{\textbf{An illustration of the different partitioning techniques.}}
	\label{fig:overview}
	\afterCaptionSpacing{}

\end{figure*}

\label{sec:impl}
In this section, we first describe the spatial indexing techniques that we implemented in our work (Section~\ref{subsec:techniques}). We then proceed to explain how we built the learned indexes (Section~\ref{subsec:index_building}). Finally, Sections~\ref{subsec:range_query}-~\ref{subsec:join_query} explain how we implemented each of the queries covered in this work (i.e., range, point, distance, and spatial join).

\subsection{Indexing Techniques}\label{subsec:techniques}

Multidimensional access methods are broadly classified into two categories: Point Access Methods (PAMs) and Spatial Access Methods (SAMs)~\cite{mdaccessmethods}.
PAMs are designed to handle point data without spatial extent.
SAMs, however, manage extended objects such as linestrings and polygons. 
In this work, we focus mainly on PAMs.

\emph{Spatial partitioning} is the process of splitting a spatial dataset into partitions, or cells, where objects within the same partition are close to each other in space. Spatial partitioning techniques can be divided into two categories: space partitioning ones, which partition the embedded space, and data partitioning ones, which partition the data space.
In this work, we employ three space partitioning techniques: linearization using the Hilbert curve, fixed-grid~\cite{fixedgrid}, and Quadtree~\cite{quadtree}. 
Additionally, we employ three data partitioning techniques: adaptive-grid~\cite{gridfile}, K-d tree~\cite{kdtree}, and Sort-Tile-Recursive (STR)~\cite{str_packing}.

Figure~\ref{fig:overview} illustrates these techniques on a sample of the Tweets dataset that we use in our experiments (further details can be found in Section~\ref{sec:datasets}).
The figure shows the sample points and partition boundaries as dots and grid axes, respectively.

\subsubsection{Linearization using Hilbert Curve}\label{sec:hilbert_curve_linear}
In a one-dimensional data space, all data points can be sorted along a single dimension, making it easy to perform range queries by retrieving all data between given lower and upper bounds. However, applying learned models to a spatial (or multidimensional) index is challenging because multidimensional data does not have an inherent sort order. One way to tackle this challenge is to linearize the multidimensional space using space-filling curves. In this work, we utilize the Hilbert space-filling curve (SFC)~\cite{hilbert_curve} to map a multidimensional vector space to a one-dimensional space. 
As shown in Figure~\ref{fig:overview}(a), the process of linearization using the Hilbert SFC involves dividing the two-dimensional space with a uniform grid and then using the Hilbert curve to enumerate the cells of this grid. Once all cells have been enumerated, we can sort their identifiers, and thus learn an index on this sorted order. This approach is similar to the recently proposed Z-order Model (ZM) index~\cite{zm-index}, which uses the Z-curve to enumerate the cells. We chose the Hilbert curve instead as it has been shown to perform better for multi-dimensional indexing~\cite{hilbert_vs_z, hilbert_vs_z_2, hilbert_vs_z_3}. The Hilbert curve is also the choice of multidimensional clustering in the recent Databricks runtime engine~\cite{databricks_hilbert}, replacing the Z-curve.
However, we show that linearization-based techniques can suffer from skewed cases, where the queries cover a large portion of the curve, as we will demonstrate in Section~\ref{sec:range_query_performance}.
For example, if a query rectangle covers all partitions at the bottom of Figure~\ref{fig:overview}(a), the whole curve would lie within the query rectangle. This would result in scanning a large number of points that do not satisfy the query, leading to poor performance.

\subsubsection{Fixed and Adaptive Grid}
Grid-based indexing methods are primarily designed to optimize the retrieval of records from disk. They work by dividing the d-dimensional attribute space into cells, where each cell corresponds to one data page (or bucket) and stores a pointer to the data page it indexes. 
Data points that fall within the boundaries of a cell are stored on the corresponding data page.

This allows for quick navigation to the specific data page containing the desired records, rather than having to search through the entire dataset.
The fixed-grid~\cite{fixedgrid} enforces equidistant grid lines, while the adaptive-grid (or grid file~\cite{gridfile}) relaxes this constraint. 
Instead, to define the partition boundaries of the d-dimensions, the adaptive-grid introduces an auxiliary data structure containing a set of d-dimensional arrays called linear scales.
In our implementation, we divide the space along one dimension and use the other dimension as the sort dimension. 
We also note that the grid-based indexes are the \emph{simplest} to implement since they only require maintaining a vector of grid lines and computing the intersection between the vector and a given query using offset computation and binary search for fixed-grid and adaptive-grid, respectively.

\subsubsection{K-d tree}
The K-d tree~\cite{kdtree} is a binary search tree that recursively subdivides the space into equal subspaces using rectilinear (or iso-oriented) hyperplanes. 
The splitting hyperplanes, known as discriminators, alternate between the $k$ dimensions at each level of the tree.
For example, in a 2-dimensional space, the splitting hyperplanes are alternately perpendicular to the x- and y-axes. The original K-d tree partitions the \emph{space} into equal partitions.
 For example, if the input space consists of a GPS coordinate system (-90.0, -180 to 90, 180), it would be divided into equal halves (-45, -90 to 45, 90). 
This results in an unbalanced K-d tree if the data is skewed, i.e., most of the data lies in one partition.
However, we can make the K-d tree data aware by dividing the data at each level into two halves based on the median point in the data. 
This ensures that both partitions in the binary tree are balanced. 
In our work, we implemented this data-aware version of the K-d tree.

\subsubsection{Quadtree}
The Quadtree~\cite{quadtree}, along with its many variants, is a tree data structure that partitions the space similarly to the K-d tree. 
The term quadtree typically refers to the two-dimensional variant, but the concept can easily be generalized to multiple dimensions. Like the K-d tree, the Quadtree decomposes the space using rectilinear hyperplanes. 
However, it differs from the K-d tree in that it is not a binary tree.
For d dimensions, interior nodes have $2^{d}$ children.
In the case of 2 dimensions, each interior node has four children, each representing a rectangle.
The search space is recursively divided into four quadrants until the number of objects in each quadrant is below a predefined threshold, typically the page size. 
Quadtrees are generally not balanced, as the tree goes deeper in areas of higher density.

\subsubsection{Sort-Tile-Recursive packed R-tree}
An R-tree~\cite{rtree} is a hierarchical data structure that is primarily designed for the efficient execution of range queries.
The R-tree approximates arbitrary geometric objects with their minimum bounding rectangle (MBR) and stores the resulting collection of rectangles. 
Each node in the R-tree stores a maximum of N entries, each containing a rectangle $R$ and a pointer $P$.
At the leaf level, $P$ points to the actual object and $R$ is the MBR of the object.
In internal nodes, $R$ represents the MBR of the subtree pointed to by $P$.

The Sort-Tile-Recursive (STR) packing algorithm ~\cite{str_packing} is a method for filling R-trees that aims to maximize space utilization. The main idea behind STR packing is to tile the data space into an $S\times S$ grid. 
Assuming that the number of points in a data set is $P$ and the capacity of a node is $N$, $S = \sqrt{P/N}$. 
STR first sorts the data on the x-dimension (in the case of rectangles, the x-dimension of the centroid), and then divides it into $S$ vertical \emph{slices}. 
Within each vertical slice, it sorts the data on the y-dimension and packs it into nodes by grouping them into runs of length $N$, forming $S$ horizontal slices. 
This process continues recursively, resulting in completely filled nodes, except for the last node which may have fewer than $N$ elements.

\subsection{Index Building}\label{subsec:index_building}

\renewcommand{\algorithmcfname}{Algorithm}
\begin{wrapfigure}{R}{0.6\textwidth}
    \begin{minipage}{0.6\textwidth}
        \begin{algorithm2e}[H]
            \footnotesize
            \SetAlgoLined
            \DontPrintSemicolon
            \Input{$D$: the input location dataset; $l$: the partition size}
            \Output{$D'$: the partitioned and indexed input dataset}
            \BlankLine
            $D' \gets \{\}$\;
                 $P$ $\gets$ \partition{some approach from the techniques described in Section~\ref{subsec:techniques}, $l$}\;
                 \For{$p\in P$}{
                  \sort{$p$, $y$}\;
                  \buildindex{$p$, $y$}\;
                  $D' \gets D' \cup \{p\}$
                 }
                 \return $D'$
            \caption{A generic way of building learning-enhanced indexes}
            \label{building_learned_index}
        \end{algorithm2e}
    \end{minipage}
\end{wrapfigure}

In this section, we outline how we can turn the above indexing techniques into learned indexes that index a given location dataset $D$, which contains points in latitude/longitude format (referred to as the x-dimension and y-dimension respectively for ease of understanding). 
First, we partition $D$ using one of the techniques described in Section~\ref{subsec:techniques}.

Each partition has a size of $l$ points, also known as the leaf size or partition size.
Once $D$ has been partitioned, we iterate through all partitions and sort all the points within each partition on the y-dimension.
Then, we build a learned index on the y-dimension for each partition.
Algorithm~\ref{building_learned_index} outlines the index building process.

\subsection{Range Query Processing}\label{subsec:range_query}
\label{sec:query_processing}
A two-dimensional range query takes as input a query range $q$ that has a lower and an upper bound in both dimensions, represented by $(q_{xl}, q_{yl})$ and $(q_{xh}, q_{yh})$ respectively. It also takes as input a location dataset $D$, containing two-dimensional points represented by $(p_x, p_y)$. The range query returns all points in $D$ that are contained in the query range $q$.
Formally:

\begin{equation*}
\begin{aligned}[b]
  Range (q,D) = \{\; p \text{\textbar} p \in D: \; (q_{xl} \leq p_x) \wedge (q_{yl} \leq p_y) \wedge {}\\ (q_{xh} \geq p_x) \wedge (q_{yh} \geq p_y) \; \}.
\end{aligned}
\end{equation*}

\begin{wrapfigure}{R}{0.60\textwidth}
    \begin{minipage}{0.55\textwidth}
        \begin{algorithm2e}[H]
            \footnotesize
            \SetAlgoLined
            \DontPrintSemicolon
            \SetNoFillComment
            \Input{$D'$: a partitioned and indexed input dataset; $q$: a query range}
            \Output{$RQ$: a set of all points in $D'$ within $q$}
            \BlankLine
             $RQ \gets \{\}$\;
             \tcc{find intersected partitions (IP)}
             $IP$ $\gets$ \indexlookup{$D'$, $q$}\; 
             \For{$ip\in IP$}{
              \tcc{if completely inside x-dimension range}
              \uIf{$q_{xl}$ $<=$ $ip_{xl}$ {\bf and} $ip_{xh}$ $<=$ $q_{xh}$}{
                  \tcc{if also completely inside y-dimension range, copy entire partition}
                  \uIf{$q_{yl}$ $<=$ $ip_{yl}$ {\bf and} $ip_{yh}$ $<=$ $q_{yh}$}{
                      \tcc{copy all points in partition}
                      $RQ$ $\gets$ $RQ$ $\cup$ $ip$\;
                  }
                  \Else{
                      \tcc{lower bound}
                      $lb$ $\gets$ \estimatefrom{$ip$, $q_{yl}$}\;
                      \tcc{get exact lower bound}
                      $lb$ $\gets$ \localsearch{$ip$, $lb$, $q_{yl}$}\;
                      \tcc{upper bound}
                      $ub$ $\gets$ \estimateto{$ip$, $q_{yh}$}\;
                      \tcc{get exact upper bound}
                      $ub$ $\gets$ \localsearch{$ip$, $ub$, $q_{yh}$}\;
                      \tcc{copy all points between lower and upper bound}
                      $RQ$ $\gets$ $RQ$ $\cup$ $ip.$\range{$lb$, $ub$}\;
                  }
              }
              \Else{
                \tcc{lower bound}
                $lb$ $\gets$ \estimatefrom{$ip$, $q_{yl}$}\;
                $lb$ $\gets$ \searchpoint{$ip$, $lb$, $q_{yl}$}\;  
                \tcc{upper bound}
                $ub$ $\gets$ \estimateto{$ip$, $q_{yh}$}\;
                $ub$ $\gets$ \localsearch{$ip$, $ub$, $q_{yh}$}\;
                \tcc{scan}
                \For{$i \in [lb, ub]$}{
                    \tcc{$ith$ point in partition $ip$}
                    $p$ $\gets$ $ip_i$\;
                    \If{$p$ {\bf within} $q$}{
                      $RQ \gets RQ \cup \{p\}$\;
                    }
                }
              }
             }
             \return $RQ$\;
            \caption{Range Query Algorithm}
            \label{range_query_algorithm}
        \end{algorithm2e}
    \end{minipage}
\end{wrapfigure}

To accelerate query processing, we use the partitioned and indexed input dataset $D'$ generated by Algorithm~\ref{building_learned_index}. Given $D'$, range query processing works in three phases, as shown in Algorithm~\ref{range_query_algorithm}:

\noindent \textbf{Phase I: Index Lookup.} The index lookup phase involves identifying the partitions that intersect with the given range query using the index directory, i.e., the grid directories or trees. These partitions are represented by $IP$, which stands for intersected partitions and is reflected in line 2 of Algorithm~\ref{range_query_algorithm}. Note that the specific method used for this step depends on the partitioning technique.

\noindent \textbf{Phase II: Boundary Refinement.} After identifying the intersected partitions in the index lookup phase, the next step is to locate the bounds of the query on the sorted dimension within each partition. 
However, when a partition is fully contained within the query range, there is no need for boundary refinement, and all points within that partition can be immediately returned. This is reflected in line 6 of Algorithm~\ref{range_query_algorithm}. 
Otherwise, when the query only partially intersects with a partition, there are two cases:
(1) the partition is fully inside the x-dimension range. In this case, we employ a search technique to compute both the lower and upper bounds, and then copy all points within these bounds (reflected in lines 8-12 of the algorithm), 
(2) the partition is not fully contained within the x-dimension range. In this case, we compute the lower and upper bounds on the sorted y-dimension and then switch to the scan phase.

Typically, binary search is used as the search technique.
In this paper, we propose replacing binary search with a learned model.
Specifically, we use the RadixSpline index~\cite{radixspline, radixspline2} to efficiently search the sorted dimension of our data. RadixSpline consists of two components: a set of spline points and a radix table. The radix table is used to quickly identify the spline points for a given lookup key (in our case, the dimension over which the data is sorted). At lookup time, the radix table is consulted first to determine the range of spline points to examine.  Next, these spline points are searched to locate the spline points surrounding the lookup key. Finally, linear interpolation is applied to predict the position of the lookup key within the sorted array.
 
Given the inherent error introduced by the RadixSpline (and generally, learned indexes), a local search (referred to as $LocalSearch()$ in Algorithm~\ref{range_query_algorithm}) is necessary to find the exact lookup point, which, in our context, corresponds to the query bound.
Without loss of generality, we describe the local search procedure for the computation of the lower query bound.
For range scans, there are two possible scenarios. 
In the first scenario, the estimated value from the spline is lower than the true lower bound within the sorted dimension. 
Thus, we scan the partition upward until we reach the lower bound. 
Conversely, when the estimated value is higher than the actual lower bound, we scan the partition downward until we reach the lower bound while also materializing all encountered points.
Consequently, in this scenario, the local search does not incur additional materialization costs (unless the estimated value exceeds the query's upper bound), as the points within the query bounds are materialized in any case.

\noindent \textbf{Phase III: Scan.} When the partition partially intersects the x-dimension range, then after determining the bounds of the query on the sorted dimension in the boundary refinement phase, the final step is to scan the partition to find the qualifying points on the x-dimension. 
During this scan phase, we iterate through the partition starting from the determined lower bound and continue until we reach either the upper bound of the query on the sorted y-dimension or the end of the partition. This process is reflected in Algorithm~\ref{range_query_algorithm} from line 14 onwards.

\subsection{Point Query Processing}\label{subsec:point_query}

In this study, we also implement point queries, in keeping with the trend in recent research~\cite{zm-index, effectively_learning_spatial_indices, efficiently_learning_spatial_indices}. A point query takes as input a query point $q_p$, and a set of geometric objects $D$.
The query returns true if $q_p$ is found within $D$, and false if it is not.
Formally:

\begin{equation*}
\begin{aligned}[b]
  Point (q_p,D) = \ \exists\; p \in D.\ q_p.x = p.x
  \wedge q_p.y = p.y.
\end{aligned}
\end{equation*}

\begin{wrapfigure}{R}{0.5\textwidth}
    \begin{minipage}{0.5\textwidth}
        \begin{algorithm2e}[H]
            \footnotesize
            \SetAlgoLined
            \DontPrintSemicolon
            \SetNoFillComment
            \Input{$D'$: a partitioned and indexed input dataset; $q_p$: a query point}
            \Output{$true$ if the point $q_p$ is in $D'$, $false$ otherwise}
            \BlankLine
            \tcc{find intersected partition (IP)}
            $IP$ $\gets$ \indexlookup{$D'$, $q_p$}\;
            \tcc{search within the partition}
            \uIf{$IP$ $\neq$ $\emptyset$}{
                \tcc{get estimate}
                $est$ $\gets$ \estimatefrom{$IP$, $q_p.y$}\;
                $found$ $\gets$ \searchpoint{$ip$, $est$, $q_p$}\;
                \return $found$
            }
            \Else{\return $false$}
            \caption{Point Query Algorithm}
            \label{point_query_algorithm}
        \end{algorithm2e}
    \end{minipage}
\end{wrapfigure}

We use the partitioned and indexed dataset, $D'$, from Algorithm~\ref{building_learned_index}, to speed up point query processing. 
Point query processing is outlined in Algorithm~\ref{point_query_algorithm}. 
First, we issue an $IndexLookup()$ using a degenerate rectangle from the query point $q_p$. 
Since the point can only exist in one partition, we first check if the $IndexLookup()$ phase produces any intersected partition. 
If no intersected partition is found, we immediately return false. 
Next, we search for the point within the intersected partition using a two-step process: (i) we estimate the location of the point in the y-dimension of the partition using the learned search technique, and (ii) we refine the result using the $SearchPoint()$ procedure, which aims to mitigate the error introduced by the learned search technique, similarly to the $LocalSearch()$ procedure used in range query processing.

For the RadixSpline, there can now be three cases for the $SearchPoint()$ procedure. 
First, if the estimated value from the spline is lower than the actual lower bound on the sorted dimension, we simply scan upward comparing the elements on the sorted dimension until we reach the actual lower bound.
We then continue scanning on both dimensions until we find the query point or reach the upper bound of the partition.
Note that for two points to be considered equal, their values should match on both dimensions. 
Second, if the estimated value is higher than the upper bound on the sorted dimension, we scan downward comparing the values on the sorted dimension until we reach the upper bound. 
We then continue scanning downward and comparing on both dimensions until we find the query point or reach the lower bound of the partition. 
Third,  if the estimated value matches the query point's value on the search dimension, we perform an upward scan, comparing on both dimensions, to locate the query point. If the partition's upper bound is reached without finding the point, we then continue with a downward scan, again comparing on both dimensions, until we locate the query point or reach the partition's lower bound.

In the case of a binary search, the process is simpler. The value of the query point on the sorted dimension is used to find the lower bound. Then, we scan upward, comparing on both dimensions until the query point is found.

\subsection{Distance Query Processing}\label{subsec:distance_query}

A distance query takes a query point $q_p$, a distance $d$, and a set of geometric objects $D$. It returns all objects in $D$ that lie within the distance $d$ of query point $q_p$. 
Formally:
\begin{equation*}
  Distance (q_p,d,D) = \{\; p\text{\textbar}p \in D \; \wedge \; \mathrm{dist}(q_p,p) \leq d \}.
\end{equation*}

\begin{wrapfigure}{R}{0.50\textwidth}
    \begin{minipage}{0.50\textwidth}
        \begin{algorithm2e}[H]
            \footnotesize
            \SetAlgoLined
            \DontPrintSemicolon
            \SetNoFillComment
            \Input{$D'$: partitioned and indexed input dataset; $q_p$: a query point; $d$: distance}
            \Output{$DQ$: a set of all points in $D'$ within distance $d$ of $q_p$}
            \BlankLine
                $DQ \gets \{\}$\;
                \tcc{Get minimum bounding rectangle (mbr) of the circle}
                $mbr$ $\gets$ \getmbr{$q_p$, $d$}\;
                \tcc{Filter using $mbr$}
                $RQ$ $\gets$ \rangequery{$D'$, $mbr$}\;
                \tcc{Refine}
                \For{$p \in RQ$}{
                    \If{\withindistance{$p$, $q_p$, $d$}}{
                        $DQ \gets DQ \cup \{p\}$\;
                    }
                }
            \return $DQ$\;
            \caption{Distance Query Algorithm}
            \label{distance_query_algorithm}
        \end{algorithm2e}
    \end{minipage}
  \end{wrapfigure}

As in the case of Range query processing (Section~\ref{sec:query_processing}), we use the partitioned and indexed input dataset $D'$ from Algorithm~\ref{building_learned_index} for faster query processing.
The implementation of the distance query employs the \emph{filter and refine}~\cite{filterrefine} approach, which is commonly used in popular database systems such as Oracle Spatial~\cite{quadtree_and_rtree_oracle}. Note that using the \emph{filter and refine} approach is also prevalent in many \emph{recent} research works where various spatial queries (e.g., kNN queries, distance queries, and spatial join queries) are first decomposed to range queries as a preliminary filter, followed by query-specific refinement~\cite{effectively_learning_spatial_indices, rlr_tree, piecewise_sfcs, lisa-sigmod}.

Algorithm~\ref{distance_query_algorithm} shows the algorithm for distance query processing.
We first filter using a rectangle (reflected in line 1 of Algorithm~\ref{distance_query_algorithm}), whose corner vertices are at a distance of \emph{d} from the query point \emph{q}.
We issue a range query using this rectangle, and then refine the resulting candidate set of points using a \emph{withinDistance} predicate.
Note that we are using GPS coordinates (i.e., a Geographic coordinate system). Therefore, special attention must be given if either of the poles or the 180th meridian is within the query distance $d$.
We compute the coordinates of the minimum bounding rectangle by moving along the geodesic arc as described in~\cite{bronshtein2013handbook} and then handle the edge cases of the poles and the 180th meridian. 

However, currently, we only use one bounding box.
This approach is not optimal as it can result in materializing a large number of unnecessary points when the 180th meridian falls within the query distance.
One way to improve the efficiency is to break the bounding box into two parts, one on either side of the 180th meridian. We leave this optimization for future work.
After materializing all the points within the MBR of $q$ and $d$, we refine this candidate set of points. 
To that end, we compute the Haversine distance between the query point $q$ and each of the candidate points and add to the final result all the points that are found to be within the specified distance $d$.

\subsection{Join Query Processing}\label{subsec:join_query}

A spatial join combines two input spatial datasets, $R$ and $S$, using a specified join predicate $\theta$ (such as overlap, intersect, contains, within, or withindistance). It returns a set of pairs $\left( r,s \right)$ where $r \in R$, $s \in S$ that meet the join predicate $\theta$.

Formally:
\begin{equation*}
   R \bowtie_\theta S = \{\; (r,s)\; \text{\textbar}\; r \in R,\; s \in S,\; \theta (r,s)\;  \mathrm{holds} \; \}.
\end{equation*}

\begin{wrapfigure}{R}{0.6\textwidth}
    \begin{minipage}{0.6\textwidth}
        \begin{algorithm2e}[H]
            \footnotesize
            \SetAlgoLined
            \DontPrintSemicolon
            \SetNoFillComment
            \Input{$D'$: partitioned and indexed input dataset; $polygons$: a set of polygons}
            \Output{$JQ$: a set of sets, a set of points within each polygon in $polygons$}
            \BlankLine
            $JQ \gets \{\}$\;
            \For{$polygon \in polygons$}{
              \tcc{Get minimum bounding rectangle (mbr) of the polygon}
              $mbr$ $\gets$ \getmbr{$polygon$}\;
              \tcc{Filter using $MBR$}
              $RQ$ $\gets$ \rangequery{$D$, $mbr$}\;
              $contained \gets \{\}$\;
              \tcc{Refine}
              \For{$p \in RQ$}{
                \If{\contains{$polygon$, $p$}}{
                  $contained \gets contained \cup \{p\}$
                }
              }
              $JQ \gets JQ \cup \{contained\}$
            }
            \return $JQ$\;
            \caption{Join Query  Algorithm}
            \label{join_query_algorithm}
        \end{algorithm2e}
    \end{minipage}
  \end{wrapfigure}

We implemented a join query between a set of polygons and the partitioned and indexed input location dataset $D'$. 
The join algorithm is outlined in Algorithm~\ref{join_query_algorithm} and is based on the \emph{filter and refine}~\cite{filterrefine} approach.


This involves using the minimum bounding rectangle of each polygon to perform a range query.
We then refine the candidate set of points using \emph{contains} as the predicate $\theta$, thus computing all points contained in each polygon. 
We implemented the contains predicate using the ray-casting algorithm, where a ray is casted from the candidate point to a point outside the polygon, and then the number of intersections with polygon edges is counted. 
Some polygons could potentially contain hundreds or thousands of edges. 
Therefore, to facilitate a quick lookup of the edges intersected with the ray, we index the polygon edges in an interval tree. 
We implemented the interval tree using a binary search tree.
\section{Evaluation}
\label{sec:evaluation}

\subsection{Experimental Setup}\label{sec:hardware}
\textbf{Hardware Configuration.} All experiments were performed single-threaded on an Ubuntu 18.04 machine equipped with an Intel Xeon E5-2660 v2 CPU (2.20\,GHz, 10 cores, 3.00\,GHz turbo)\footnote{CPU: \url{https://ark.intel.com/content/www/us/en/ark/products/75272/intel-xeon-processor-e5-2660-v2-25m-cache-2-20-ghz.html}} and 256\,GB DDR3 RAM. We use the \emph{numactl} command to bind the thread and memory to one node to avoid NUMA effects. CPU scaling was also disabled during benchmarking using the \emph{cpupower} command.
\\
\textbf{Software Configuration.}
In all our experiments, we sort on the longitude value of the location within each partition.
The currently available open-source implementation of RadixSpline only supports integer values. However, most spatial datasets contain floating-point values. To address this issue, we adapted the RadixSpline implementation to work with floating-point values. 
We set the spline error to 32 for all experiments in our RadixSpline implementation.

\begin{figure}[t]
\centering
  \begin{subfigure}[t]{.31\linewidth}
  \centering
    \resizebox{\textwidth}{!}{\fbox{\includegraphics{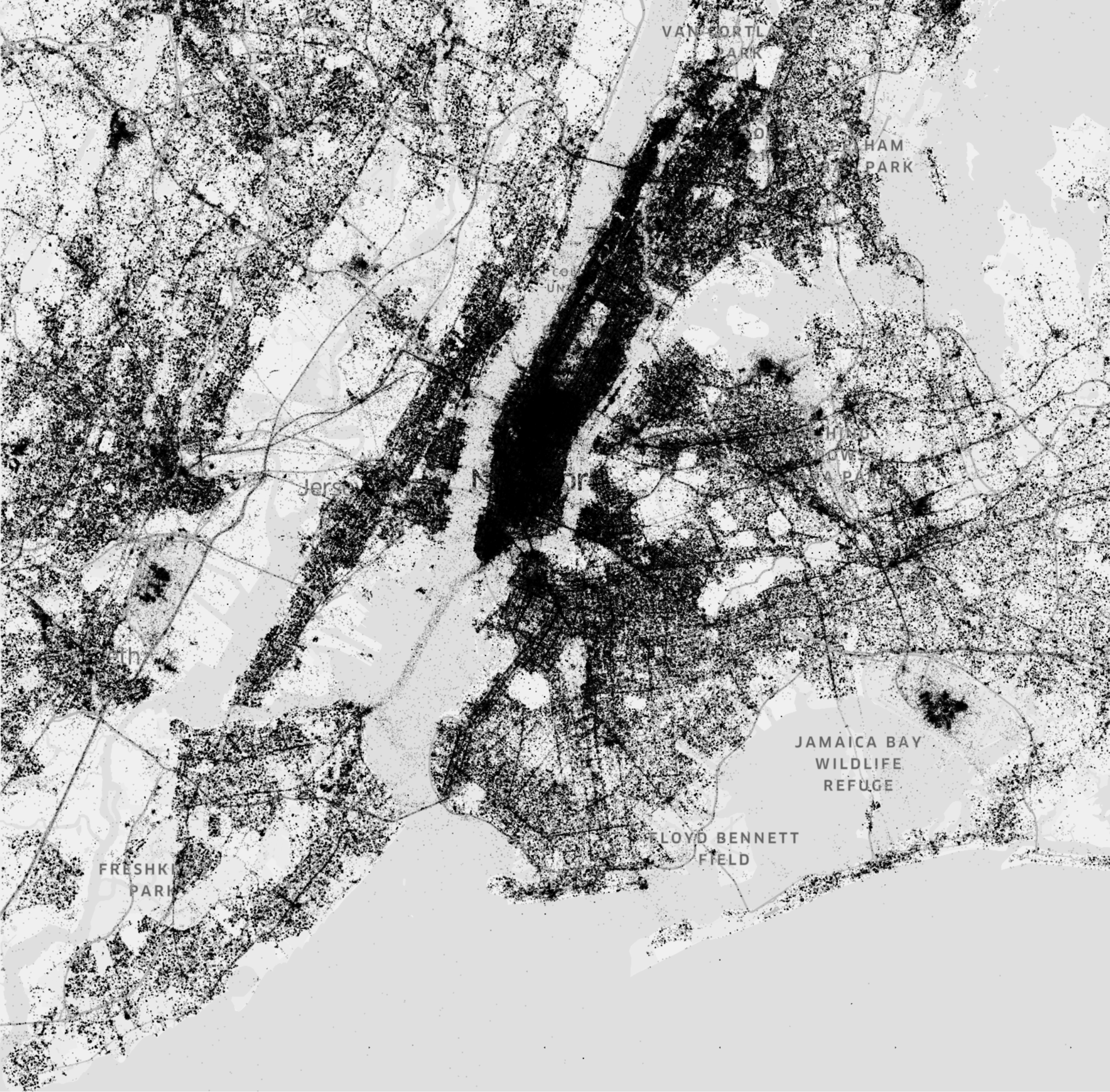}}}
    \caption{Twitter}
    \label{fig:tweets_dataset_bw}
  \end{subfigure}
\hfill
\begin{subfigure}[t]{.31\linewidth}
  \centering
    \resizebox{\textwidth}{!}{\fbox{\includegraphics{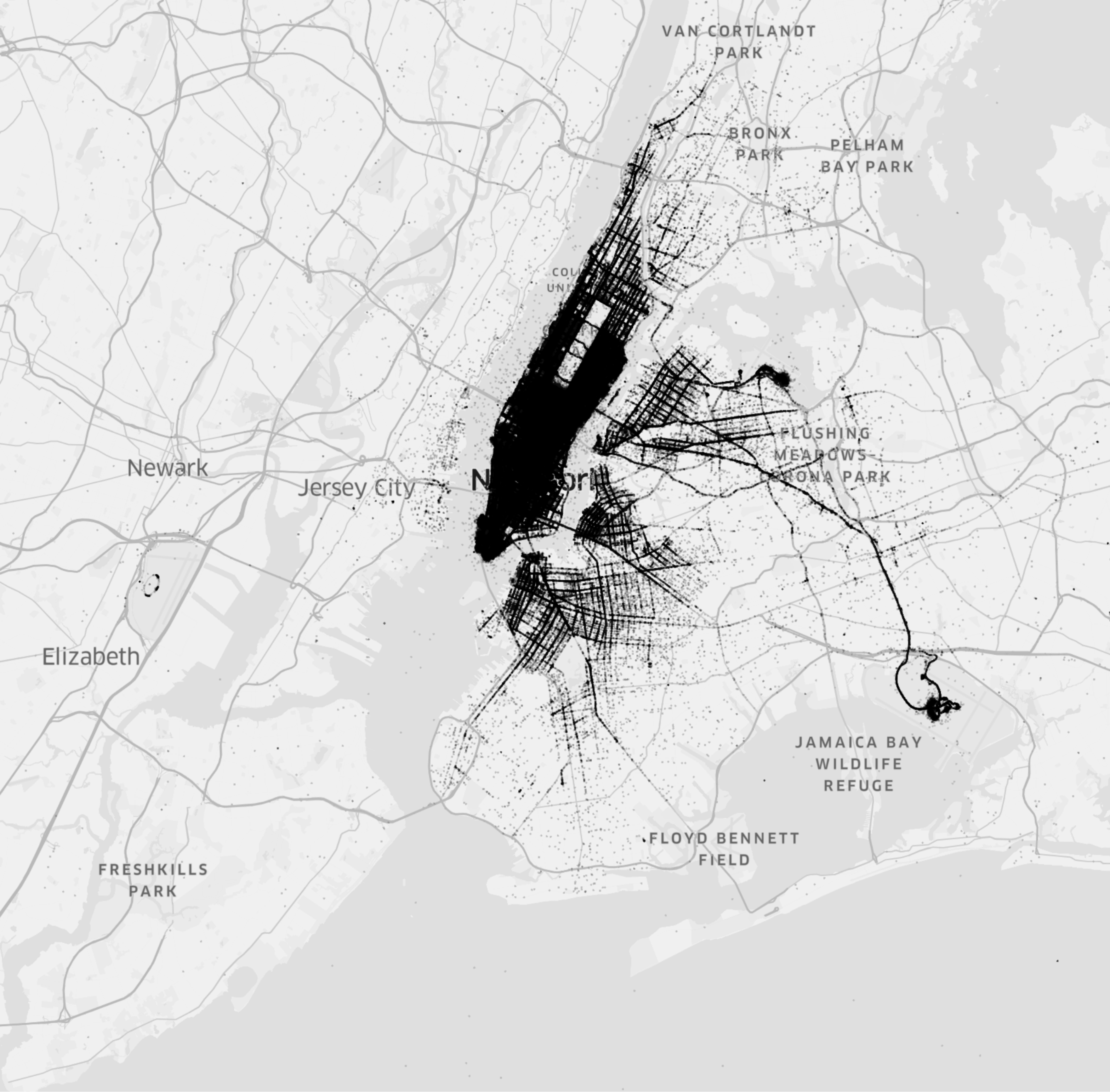}}}
    \caption{\protect\raggedright Taxi Trips}
    \label{fig:taxi_dataset_bw}
  \end{subfigure}
\hfill
  \begin{subfigure}[t]{.32\linewidth}
  \centering
    \resizebox{\textwidth}{!}{\fbox{\includegraphics{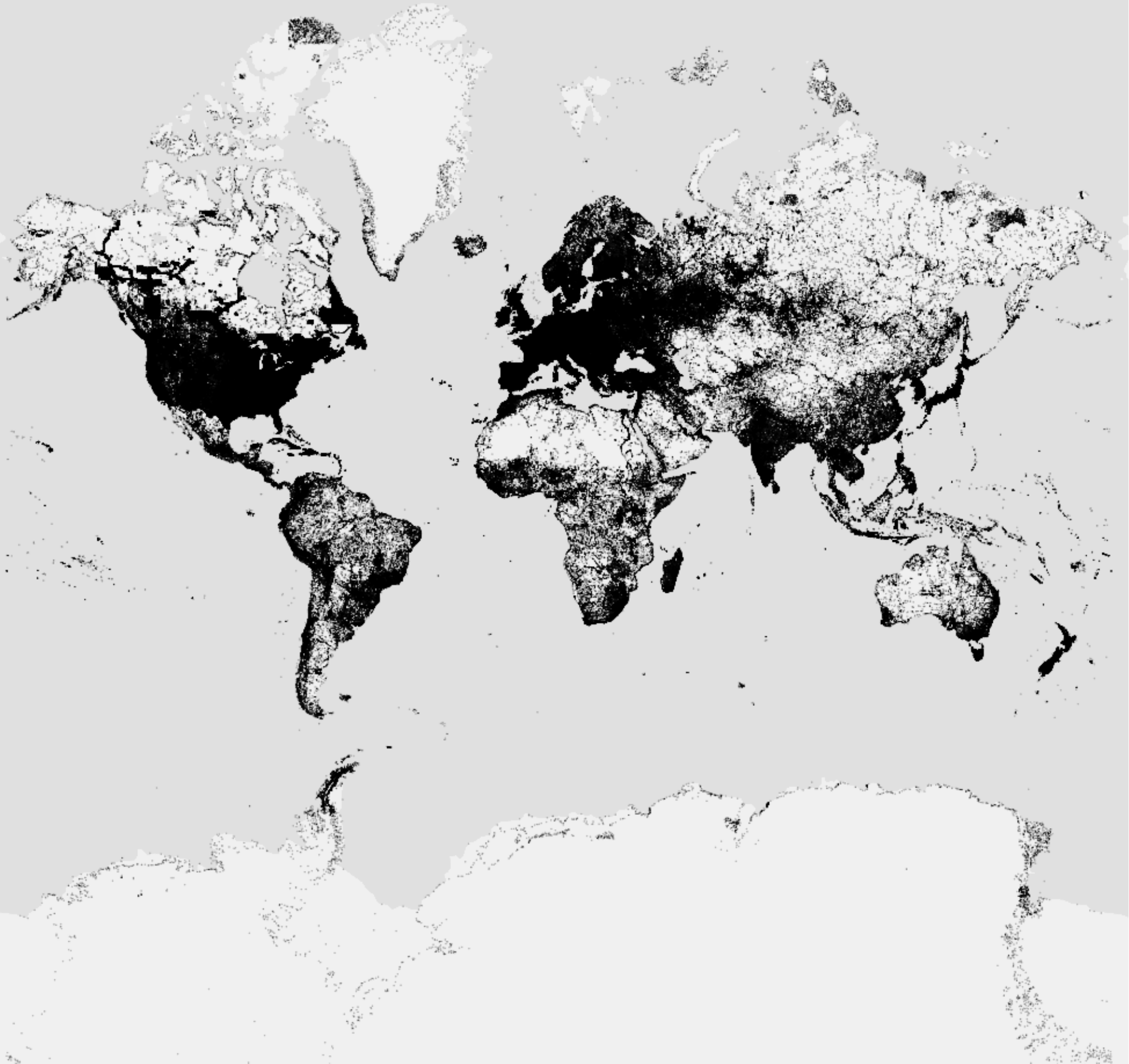}}}
    \caption{OSM}
    \label{fig:osm_dataset_bw}
  \end{subfigure}
\caption{Datasets: (a) Tweets are spread across New York, (b) NYC Taxi trips are clustered in central New York, and (c) All Nodes dataset from OSM.}
\label{fig:datasets_bw}
\end{figure}

\subsection{Datasets and Queries}\label{sec:datasets}

For evaluation, we used three datasets, the New York City Taxi Rides dataset~\cite{nyctaxidata} (NYC Taxi Rides), geo-tagged tweets in the New York City area (NYC Tweets), and Open Streets Maps (OSM). NYC Taxi Rides contains 305 million taxi rides from 2014 and 2015. NYC Tweets data was collected using Twitter's Developer API~\cite{nyctweets} and contains 83 million tweets. The OSM dataset is taken from~\cite{howgood} and contains 200M records from the All Nodes (Points) dataset. 
Figure~\ref{fig:datasets_bw} shows the spatial distribution of the three datasets.
We further generated two types of query workloads for each of the three datasets: skewed queries, which follow the distribution of the underlying data, and uniform queries. 
For each type of query workload, we generated six different workloads ranging from 0.00001\% to 1.0\% selectivity. For example, in the case of the Taxi Rides dataset (305M records), these queries would materialize from 30 to 3 million records. The query workloads consist of one million queries each. To generate skewed queries, we select a record from the data and expand its boundaries (using a random ratio in both dimensions) until the selectivity requirement of the query is met. 
For uniform queries, we generate points uniformly in the embedding space of the dataset and expand the boundaries similarly until the selectivity requirement of the query is met. The query selectivity and the type of query are mostly application-dependent. For example, consider a user issuing a query to find a popular pizzeria nearby on Google Maps. The expected output for this query should be a handful of records, i.e., the query selectivity is low (a list of 20-30 restaurants near the user). 
On the other hand, a query on an analytical system would materialize many more records (e.g., find the average cost of all taxi rides originating in Manhattan).

\begin{figure*}[t]
    \includegraphics[width=1.00\textwidth,height=0.5\textheight,keepaspectratio]{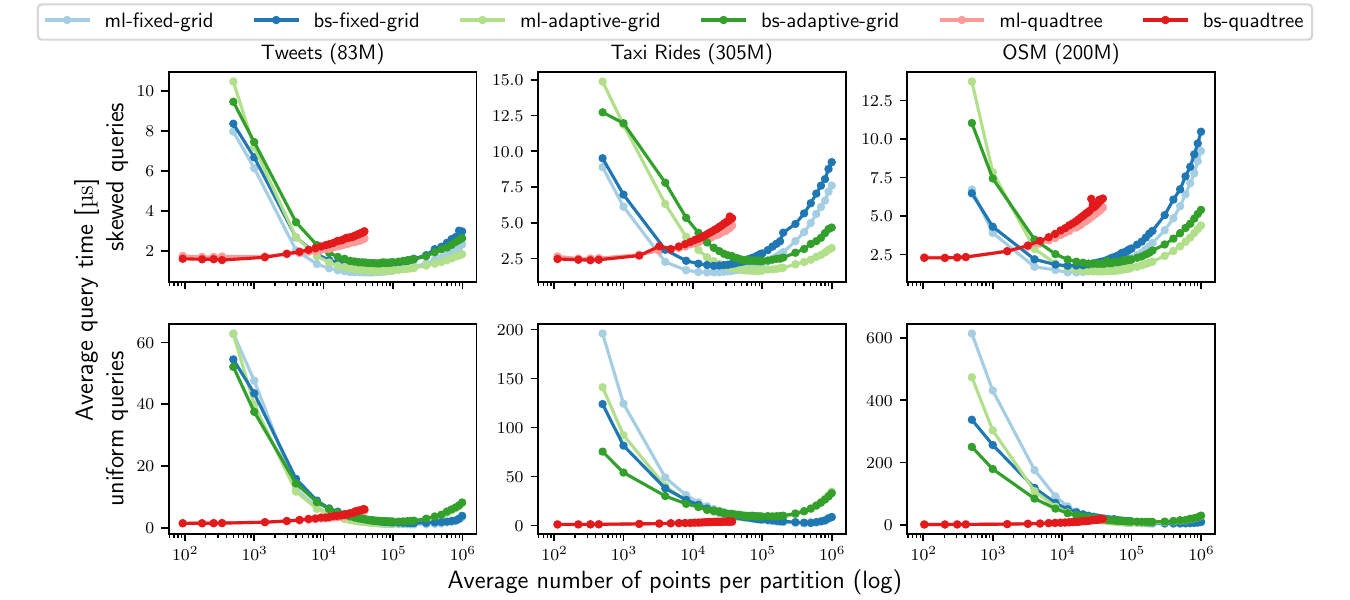}
    \caption{Range Query Configuration - ML vs. BS for low selectivity (0.00001\%).}
    \label{fig:learning_vs_bs_low}
\end{figure*}

\subsection{Baselines}\label{sec:baselines}
Firstly, we compare the performance of learned indexes and binary search as search techniques within a partition. Furthermore, we compare our implementation of the learned indexes with the two best-performing indexes from earlier studies~\cite{spatiallibs, fullspatiallibs}, which compared state-of-the-art spatial libraries. 
More specifically, for range and distance queries, we compare our implementation with the STRtree implementation from the Java Topology Suite (JTS) and the S2PointIndex from Google S2. 
For join queries, we use the S2ShapeIndex provided by Google S2. 
The source code used in this work is available on GitHub\footnote{\url{https://github.com/varpande/learnedspatial}}. Additionally, we will open source the implementation of the learned indexes and the query workloads with the camera-ready version of this work. 
Given the popularity of machine-learned indexes in current research, we believe that our implementations will be useful in evaluating many influential works to come in the near future.

\subsection{Range Query Performance}\label{sec:range_query_perf}
In this section, we first explore the tuning of partition sizes and why the tuning is crucial to obtain optimal performance. Next, we present the total query runtime when the partition size for each index is tuned for optimal performance. 

\subsubsection{Tuning Indexing Techniques}\label{sec:range_query_tuning}

Recent work in learned multidimensional and spatial indexes has focused on learning from the data and the query workload. The essential idea behind learning from both the data and the query workload is that a particular use case can be instance-optimized~\cite{instance_optimized_1, instance_optimized_2}. To study this effect, we conducted multiple experiments on the three datasets by varying the sizes of the partitions, tuning them on two workloads with different selectivities (to cover a broad spectrum, we tune the indexes using queries with low and high selectivity) for both skewed and uniform queries. 
We omit the results for tuning the indexing techniques for the rest of the queries (point, distance, and join queries) as they are similar to the ones for range queries.

\begin{figure*}[t]
    \centering
    \begin{minipage}{0.48\linewidth}
        \centering
        \includegraphics[width=\linewidth]{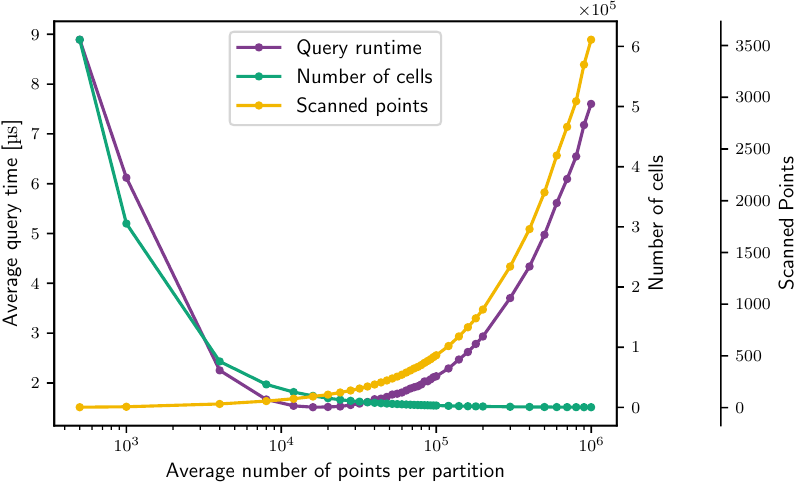}
        \caption{Effect of the number of cells and scanned points for fixed-grid on Taxi Trip dataset for skewed queries (0.00001\% selectivity).}
        \label{fig:effect_of_cells}
    \end{minipage}\hfill
    \begin{minipage}{0.50\linewidth}
        \centering
        \includegraphics[width=\linewidth]{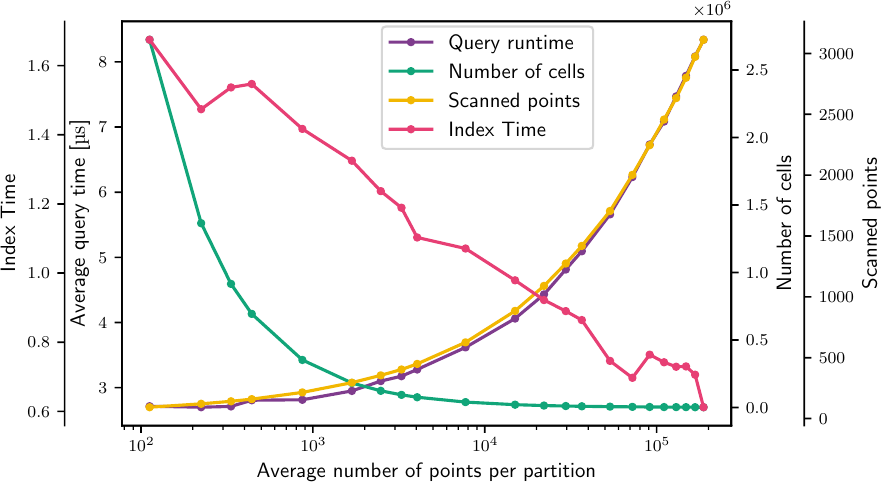}
        \caption{Effect of the number of cells and scanned points for Quadtree on Taxi Trip dataset for skewed queries (0.00001\% selectivity).}
        \label{fig:effect_of_cells_quadtree}
    \end{minipage}
\end{figure*}

\begin{table*}
\scriptsize
\resizebox{\textwidth}{!}{%
\begin{tabular}{lrrrrrrrrrrrrrrrrrrrr}
\toprule
{} & \multicolumn{6}{c}{Taxi Trips (Skewed Queries)} & \multicolumn{6}{c}{Taxi Trips (Uniform Queries)} \\
\midrule
{} & \multicolumn{2}{c}{Fixed} & \multicolumn{2}{c}{Adaptive} & \multicolumn{2}{c}{Quadtree} & \multicolumn{2}{c}{Fixed} & \multicolumn{2}{c}{Adaptive} & \multicolumn{2}{c}{Quadtree} \\
\midrule
{Selectivity (\%)}       &  ML       &    BS      &  ML       &  BS       &  ML        &  BS       &  ML        &  BS       & ML       &  BS       &    ML       &    BS   \\
\midrule
0.00001             &  1.78    &   2.35    &  1.86    &  2.40    &  2.77     &  2.51    &  2.02     &  2.58    & 81.4    &  10.54   &  1.48      &  1.31  \\
0.0001              &  4.54    &   5.82    &  4.67    &  6.12    &  6.12     &  5.82    &  5.85     &  6.91    & 228.1   &  27.69   &  3.69      &  3.42   \\
0.001               &  14.97   &   18.83   &  15.32   &  19.49   &  20.84    &  19.47   &  22.87    &  24.34   & 708.8   &  87.49   &  13.59     &  12.98  \\
0.01                &  90.13   &   97.04   &  89.48   &  95.96   &  117.01   &  104.37  &  141.24   &  151.47  & 2634.4  &  309.62  &  98.85     &  112.77 \\
0.1                 &  678.12  &   698.39  &  675.14  &  696.49  &  922.67   &  793.96  &  988.35   &  922.96  & 9609.9  &  1174.79 &  891.24    &  1101.95 \\
1.0                 &  8333.94 &   8408.15 &  8301.56 &  8399.69 &  10678.04 &  9512.29 &  8843.71  &  8753.68 & 8574.84 &  8836.28 &  10647.97  &  12377.14  \\
\bottomrule
\end{tabular}
}
\centering
\caption{Total query runtime (in microseconds) for both RadixSpline (ML) and binary search (BS) for Taxi Rides dataset on skewed and uniform query workloads (parameters are tuned for selectivity 0.00001\%).}
\label{tab:runtime}
\end{table*}

Figure~\ref{fig:learning_vs_bs_low} shows the effect of tuning when the indexes are tuned for the lowest selectivity workload for the two query types. It can be seen in the figure that it is essential to tune the grid indexing techniques for a particular workload. Firstly, they are \emph{highly} susceptible to the size of the partition. As the size of the partition increases, we notice an improvement in the performance until a particular partition size is reached that corresponds to the optimal performance. After this point, increasing the size of the partitions only degrades the performance. This shows that determining the optimal partition is crucial for optimal performance. For example, fixing the partition size of fixed-grid to 100 points per partition (a fairly common default value of the leaf size for an index in many open-source spatial libraries) results in up to \emph{300$\times$ worse} performance compared to the optimal partition size.
It can be seen that, usually, for grid (single-dimension) indexing techniques, the optimal partition sizes are much larger compared to indexing techniques that filter on both dimensions (only Quadtree is shown in the figure but the same holds for the other indexing techniques we have covered in this work; we do not show the other trees because the curve is similar for them). 
Due to the large partition sizes in grid indexing techniques, we notice a large increase in performance while using a learned index compared to binary search. This is especially evident for skewed queries (which follow the underlying data distribution). We encountered a speedup from 11.79\% to 39.51\% compared to binary search. Even when we tuned a learned index to a partition size that corresponds to the optimal performance for binary search, we found that the learned index frequently outperformed the binary search. 
Learned indexes do not help much for indexing techniques that filter on both dimensions.
In contrast, as shown in Table~\ref{tab:runtime}, the performance of Quadtree (and STRtree) dropped in many cases.
The reason is that the optimal partition size for these techniques is very low (less than 1,000 points per partition for most configurations). The refinement cost for the learned indexes is an overhead in such cases. The k-d tree, on the other hand, contains more points per partition (from 1200 to 7400) for the optimal configuration for Taxi Trips and OSM datasets, and thus the learned indexes perform faster by 2.43\% to 9.17\% than binary search. For the Twitter dataset, the optimal configuration contains less than 1200 points per partition, and we observed a similar drop in performance using learned indexes.

\begin{figure*}[t]
    \includegraphics[width=1.00\textwidth,height=0.5\textheight,keepaspectratio]{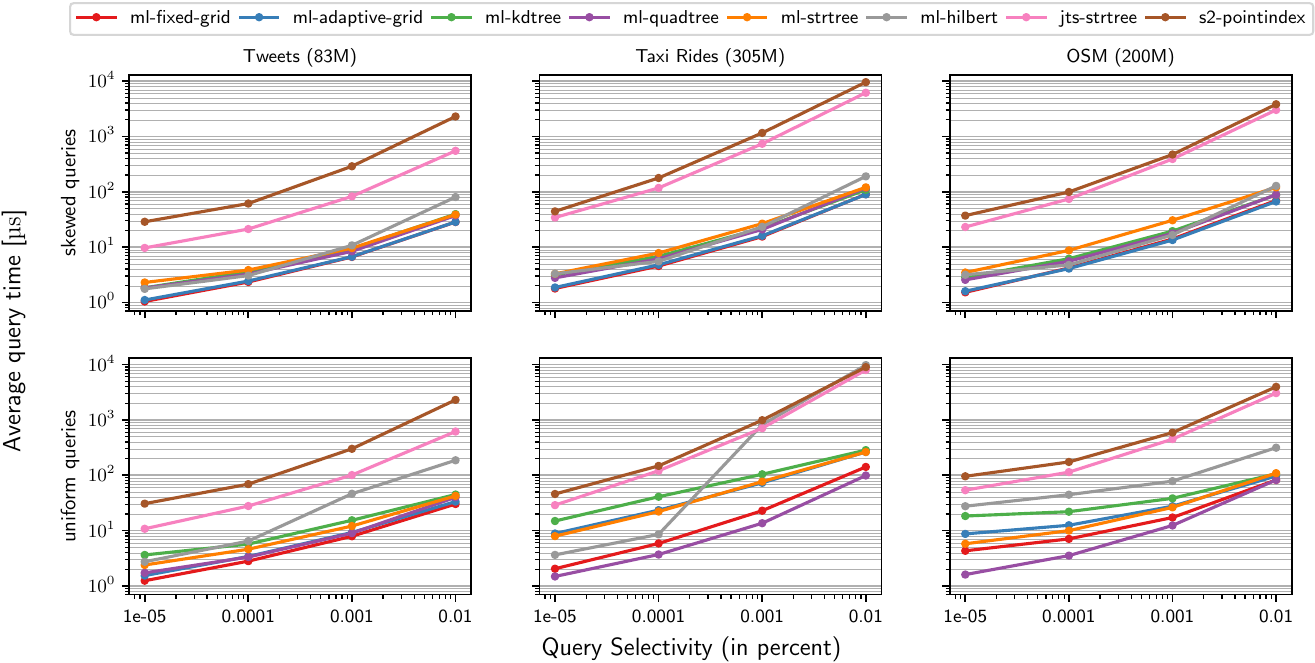}
    \caption{\textbf{Range Query Runtime - }\textrm{Total range query runtime on skewed and uniform queries for the three datasets.}}
    \label{fig:runtime_low}
\end{figure*}

Figure~\ref{fig:effect_of_cells} shows the effect of the number of cells and the number of points that are scanned in each partition on the query runtime for fixed-grid on Taxi Trips dataset for the lowest selectivity. As the number of points per partition increases (i.e., there are fewer partitions), the number of cells decreases. At the same time, the number of points that need to be scanned for the query increases. The point where these curves meet is the optimal configuration for the workload, which corresponds to the lowest query runtime. For tree structures, the effect is different. Figure~\ref{fig:effect_of_cells_quadtree} shows that the structures that filter on both dimensions do most of the pruning in the index lookup. The dominating cost in these structures is the number of points scanned within the partition, and the query runtime is directly proportional to this number. To minimize the number of points scanned, they do most of the pruning during index lookup, which requires more partitions (i.e., fewer points per partition). Tree-based structures pay more during the index lookup phase, which requires chasing pointers (random access) during the index lookup and leads to more cache misses. However, once the desired partition is reached, these index structures scan very few points, as most of the partitions qualify for the query.

\greybox{\textbf{Key Takeaways.} Tuning partition sizes is crucial for grid-based indexing techniques. The optimal size of grid partitions is typically large, which enables fast searches within the partitions using learned models. In contrast, tree-based indexes do not gain as much from learned models because optimal partition sizes are typically small.}

\subsubsection{Query Performance}\label{sec:range_query_performance}

Figure~\ref{fig:runtime_low} shows the query runtime for all learned index structures. It can be seen that fixed-grid along with adaptive-grid (1D schemes) perform the best for all cases except for uniform queries on Taxi and OSM datasets. For skewed queries, fixed-grid is 1.23$\times$ to 1.83$\times$ faster than the closest competitor, Quadtree (2D), across all datasets and selectivity. The slight difference in performance between fixed-grid and adaptive-grid comes from the index lookup. For adaptive-grid, we use binary search on the linear scales to find the first partition the query intersects with. For fixed-grid, the index lookup is almost negligible as only an offset computation is needed to find the first intersecting partition. This is also in contrast to traditional knowledge that grid-based index structures can become skewed and thus perform worse than tree-based index structures. Since the index structures and data reside in memory and are tuned to optimal partition size, the grid-based structures perform better as (1) they avoid pointer chasing as in the case of tree-based index structures, thus leading to fewer random accesses, and (2) they can utilize the \emph{fast lookups} using learned models within the large indexed partitions. This would not be possible for disk-based index structures. Note that partition sizes for optimal performance of grid-based index structures are very large for every datasets. For disk-based index structures to exhibit similar performance, it would require allocating very large pages on disk.

It can also be seen in the figure that the Quadtree is significantly better for uniform queries in the case of the Taxi Rides dataset (1.37$\times$) and OSM dataset (2.68$\times$) than the closest competitor, fixed-grid. There are two reasons for this. First, as Table~\ref{tab:stats} shows, the Quadtree intersects with fewer partitions than the other index structures. 
Second, for uniform queries, the Quadtree is more likely to traverse the sparse and low-depth region of the index. 
\begin{wraptable}{LH}{7.5cm}
\caption{Average number of partitions intersected for each indexing method for selectivity 0.00001\% on Taxi Rides and OSM datasets.}
\small
\begin{tabular}{lrrrr}
\toprule
{} & \multicolumn{2}{c}{Taxi Rides} & \multicolumn{2}{c}{OSM} \\
\midrule
{Indexing}     &  Skewed   &   Uniform  &   Skewed  &  Uniform  \\
\midrule
Fixed             &  1.97      &   7.98     &   1.72    &  23.73    \\
Adaptive          &  1.74      &   31.57    &   1.51    &  24.80    \\
k-d tree          &  1.70      &   21.62    &   1.56    &  30.95    \\
Quadtree          &  1.79      &   2.12     &   1.37    &  7.96     \\
STR               &  2.60      &   47.03    &   1.90    &  11.05    \\
\bottomrule
\end{tabular}
\centering
\label{tab:stats}
\end{wraptable}
This is consistent with previously reported findings~\cite{quadtree_uniform}, where the authors compare the Quadtree with the R*-tree and the Pyramid-Technique.

In Figure~\ref{fig:runtime_low}, we can also see the performance of the learned indexes compared to JTS STRtree and S2PointIndex. Fixed-grid is from 8.67$\times$ to 43.27$\times$ faster than the JTS STRtree. Fixed-grid is also from 24.34$\times$ to 53.34$\times$ faster than S2PointIndex. Quadtree, on the other hand, is from 6.26$\times$ to 33.99$\times$ faster than JTS STRtree, and from 17.53$\times$ to 41.91$\times$ faster than S2PointIndex.
Note that the index structures in the libraries are not tuned and are taken as is out of the box with default values. The poor performance of S2PointIndex is because it works on top of the Hilbert curve values and is not optimized for range queries. S2PointIndex and the learned linearized Hilbert curve index counterpart are rather similar in nature. Both use linearization to one dimension for indexing. The learned counterpart is learned on the sorted values of the linearized values where the underlying implementation is a densely packed array, while S2PointIndex stores these linearized values in a main-memory optimized B-tree. S2PointIndex is a B-tree on the 64-bit integers called S2CellId. T
he cell ids are a result of the Hilbert curve enumeration of a \emph{Quadtree}-like space decomposition. 
Hilbert curve (as previously mentioned in Section~\ref{sec:hilbert_curve_linear}) suffers from skewed cases where the range query rectangle covers the whole curve. 
To avoid such a case, S2PointIndex decomposes the query rectangle into four parts to reduce the overlap with the curve. 
However, it still scans many superfluous points.
\greybox{\textbf{Key Takeaways.} Fixed-grid performs the best across all queries, except for uniform queries on Taxi and OSM datasets. This is because it avoids random pointer chasing and utilizes fast learning-enhanced search within its large partitions. Quadtree outperforms fixed-grid for uniform queries on Taxi and OSM because these queries intersect with fewer partitions and only traverse the sparse, low-depth region of the index. Finally, the linearized Hilbert curve does not perform well in most cases because the queries need to scan a large portion of the curve.}

\subsection{Point Query Performance}\label{sec:point_query_performance}


Section~\ref{subsec:point_query} defines how the point query has been implemented in this work. 
JTS STRtree does not provide a way to search for a point and thus we did not implement the point query using JTS STRtree.
S2PointIndex in the Google S2 library, on the other hand, allows querying for a point.
Moreover, we run the point queries using only the skewed queries workload, which uniformly selects a random point from the dataset itself. 
This ensures that the point query actually produces a result and does not skew the results in favor of the learned indexes.
It is also consistent with real-world workloads, where it is more common to search for an existing point in the dataset, such as retrieving metadata for a specific restaurant location. 
\begin{wrapfigure}{r}{0.5\textwidth}
  \begin{center}
    \includegraphics[width=0.48\textwidth]{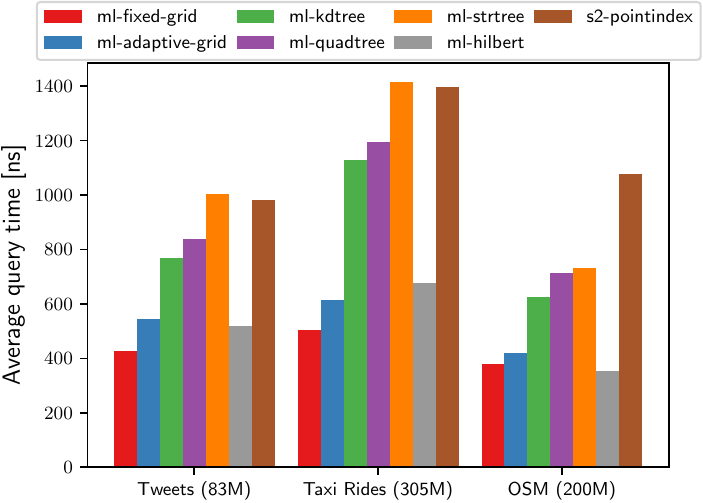}
  \end{center}
  \caption{\textbf{Point Query Performance - }\textrm{Total point query runtime for skewed queries on the three datasets.}}
  \label{fig:point_query_runtime}
\end{wrapfigure}

Figure ~\ref{fig:point_query_runtime} shows the point query runtime for all indexing techniques.
Fixed-grid is again the best-performing index for point queries on skewed workloads.
It is 1.94$\times$, 2.27$\times$, and 1.51$\times$ faster that the closest tree-based competitor, kdtree, across Tweets, Rides, and OSM datasets. However, the performance difference of fixed-grid with adaptive-grid are marginal. It is 1.37$\times$, 1.1$\times$, 1.04$\times$ faster than the adaptive-grid across the Tweets, Rides, and OSM datasets. 
Lastly, fixed-grid is also 2.44$\times$, 2.56$\times$, 2.79$\times$ faster than S2PointIndex.
Another very important observation is that the learned index on the linearized values is very competitive in the point queries.
This is counter-intuitive from the observation in range queries in Section~\ref{sec:range_query_performance}. We noted that range searches on sorted Hilbert curve values perform poorly in skewed cases where the query rectangle covers a large portion of the curve. 
However, for point queries, only one point on the curve needs to be searched, rather than scanning multiple points.
This makes the learned index on the linearized values very competitive for point queries. In fact, in the case of the OSM dataset, it is the best-performing index with an average query time of 385\emph{ns} compared to 390\emph{ns} for the fixed-grid (the best-performing index in the other two datasets).

\greybox{\textbf{Key Takeaways.} Fixed-grid performs the best across all queries. The linearized Hilbert curve is highly competitive for point queries since it only requires searching for a single point on the curve. This is in contrast to range queries where a large portion of the curve needs to be scanned.}

\subsection{Distance Query Performance}\label{sec:distance_query_performance}

\begin{figure*}[t]
    \includegraphics[width=1.00\textwidth,height=0.5\textheight,keepaspectratio]{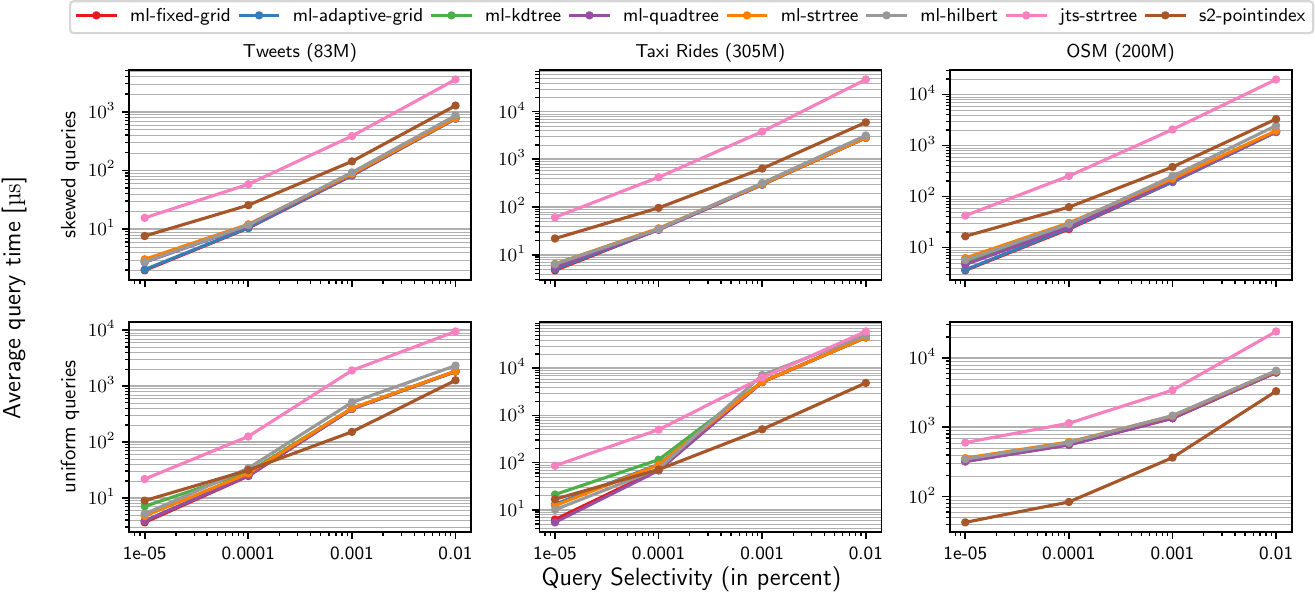}
    \caption{\textbf{Distance Query Performance - }\textrm{Total distance query runtime on skewed and uniform queries for the three datasets.}}
    \label{fig:runtime_low_distance}
\end{figure*}

We implemented the distance query using the filter and refine~\cite{filterrefine} approach (see Section~\ref{subsec:distance_query}), which is the norm in spatial databases such as Oracle Spatial~\cite{quadtree_and_rtree_oracle} and PostGIS~\cite{postgis:online}. We index GPS coordinates and use the \emph{Harvesine} distance in the refinement phase.

Figure~\ref{fig:runtime_low_distance} shows the distance query runtime for all indexing techniques as well as the two spatial indexes, S2PointIndex and JTS STRtree. We can make two important observations. First, the difference in performance between the learned indexes diminishes quickly as we increase the selectivity of the query. Grid-based indexes perform the best for lower selectivities (0.00001\% and 0.0001\%), except for uniform queries on Taxi Rides and OSM datasets where Quadtree is better, similar to range query. 
 However, as more points qualify in the filter phase, Haversine distance computation becomes the dominant cost, causing the performance of all indexing methods to converge.
Haversine distance is computationally expensive and requires multiple additions, multiplications, and divisions as well as three trigonometric function calls. Although we only use Harvesine distance on a subset of points in the filter phase, it is still expensive to compute. 

The second observation is that S2PointIndex outperforms most of the indexes for uniform queries on the OSM dataset. The reason for this is that after the filter phase, many points need refinement for uniform queries for the OSM dataset. For example, for the OSM dataset, the average number of points that need refinement after the filter phase for skewed queries is 25, 257, and 2561 for selectivities 0.00001\%, 0.0001\%, and 0.001\%, respectively. For uniform queries, the average number of points that need refinement after the filter phase is 4257, 7263, 17612 (6$\times$ to  170$\times$ more than skewed queries) for the OSM dataset. The dominant cost for most index structures is the Haversine distance computation, and thus we also do not observe much difference in performance between the learned indexes. S2PointIndex on the other hand has optimizations for distance queries, where it carefully increases the radius of the internal data structure called S2Cap (a circular disc with a center and a radius) to visit the Hilbert curve. It does not explicitly rely on the Haversine distance computation but works similarly to a range query with a radius. The S2PointIndex first searches for the point in the Hilbert curve using the query points and then increases the radius along the Hilbert curve until the query radius is satisfied. Therefore, the distance query using the S2PointIndex for uniform queries on the OSM dataset is from 1.91$\times$ to 7.75$\times$ faster than the learned indexes. This effect does not reflect in the other datasets since after the filter phase the number of points that qualify for Haversine distance computation is similar to that for the skewed queries in the OSM dataset. The comparison of learned indexes with JTS STRtree is fairer since both the learned indexes and JTS STRtree deploy the \emph{filter and refine} approach to evaluate distance queries. Fixed-grid is from 1.33$\times$ to 11.92$\times$ faster than JTS STRtree.

\greybox{\textbf{Key Takeaways.} The computation of Haversine distance dominates the cost of distance queries. While learned search offers significant performance gains for lower selectivity queries and for queries with skewed distributions, its advantages diminish for queries with higher selectivity and uniform distributions. These queries produce a large number of points for refinement (i.e., Haversine distance computations). As a result, in many cases, S2PointIndex outperforms learned indexes.}

\subsection{Join Query Performance}\label{sec:join_query_performance}

\begin{figure*}[t]
    \includegraphics[width=1.00\textwidth,height=0.5\textheight,keepaspectratio]{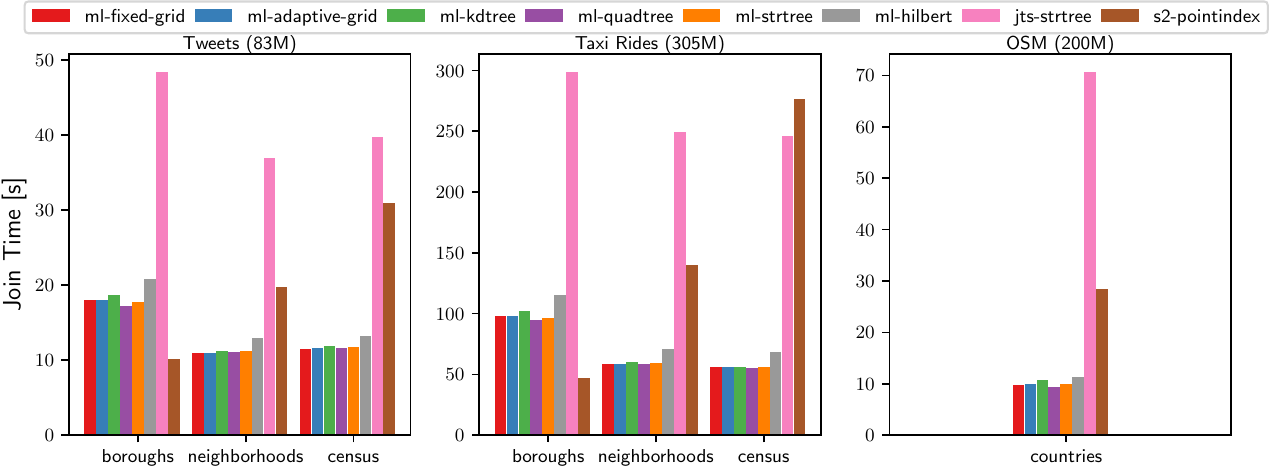}
    \caption{\textbf{Join Query Performance - }\textrm{Total join query runtime for the three datasets.}}
    \label{fig:join_query_perf}
\end{figure*}

For join queries, we utilized the \emph{filter and refine} approach for the learned indexes and JTS STRtree. We use the bounding box of the polygon objects and issue a range query on the indexed points, while in S2, we utilize the S2ShapeIndex which is specifically built to test for the containment of points in polygonal objects. As mentioned in Section~\ref{subsec:join_query}, we index the polygon objects using an interval tree in case of the learned indexes. For JTS STRtree, we utilize the PreparedGeometry\footnote{\url{https://locationtech.github.io/jts/javadoc/org/locationtech/jts/geom/prep/PreparedGeometry.html}} abstraction, to index line segments of all individual polygons, which helps in accelerating the refinement check.

We used three different polygonal datasets for the join query with the location datasets that are in the NYC area (i.e., Tweets and Taxi Rides datasets). Specifically, we used the Boroughs, Neighborhoods, and Census block boundaries consisting of five, 290, and approximately 40 thousand polygons, respectively.
For the OSM dataset, we perform the join using the Countries dataset which consists of 255 country boundaries. 
Similarly to range and distance queries, we first find the optimal partition size for each learned index and dataset.

Figure~\ref{fig:join_query_perf} shows the join query performance. It can be observed in the figure that most of the learned indexes are similar in join query performance. The reason behind this is that the \emph{filter} phase is not expensive for the join query, while the \emph{refinement} phase is the dominant cost. This result is in conformance to earlier studies~\cite{spatiallibs, fullspatiallibs}, which compared state-of-the-art spatial libraries used by hundreds of systems and other libraries. 
Although we use an interval tree to index the edges of the polygons to quickly determine the edges intersecting the ray casted from the candidate point, this phase is still expensive. 
For future work, we plan to investigate the performance using the main-memory index for polygon objects proposed in~\cite{edbtjoin}.

It can also be seen in the figure that the learned indexes are considerably faster than JTS STRtree and S2ShapeIndex for the join query. Fixed-grid, for example, is 1.81$\times$ to 2.69$\times$ faster than S2ShapeIndex and 2.7$\times$ to 3.44$\times$ faster than JTS STRtree for the Tweets dataset across all three polygonal datasets. Similarly, for the Taxi Rides dataset, fixed-grid is 2.39$\times$ to 4.96$\times$ faster than S2ShapeIndex and 3.017$\times$ to 4.49$\times$ faster than JTS STRtree. Finally, for the OSM dataset, it is 2.89$\times$ faster than S2ShapeIndex and 7.311$\times$ faster than JTS STRtree.

\greybox{\textbf{Key Takeaways.} The \emph{filter} phase of the join query produces a large number of points for the \emph{refinement} phase, which involves computationally-expensive point-in-polygon tests.
The computational overhead of the point-in-polygon tests diminishes the benefits gained from the fast \emph{filter} phase. As a result, the query performance is similar to that of the distance query, where the dominant cost is the Haversine distance computation.}

\subsection{Indexing Costs}

\begin{figure*}[t]
    \centering
    \includegraphics[width=0.8\textwidth,keepaspectratio]{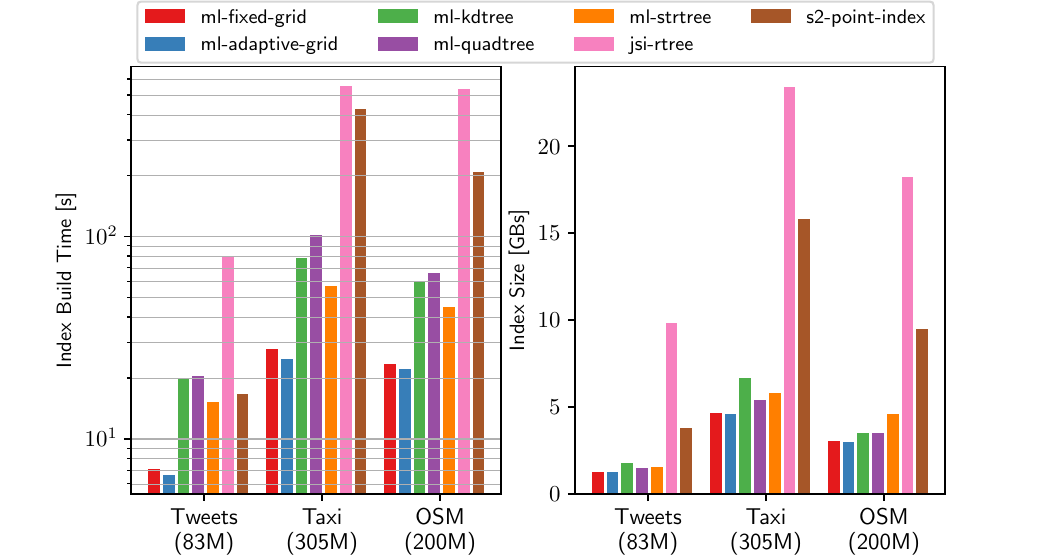}
    \caption{\textbf{Indexing Costs - }\textrm{Index build times and sizes for the three datasets.}}
    \label{fig:indexing_costs}
\end{figure*}

Figure~\ref{fig:indexing_costs} shows that fixed-grid and adaptive-grid are faster to build than tree-based learned indexes. Fixed-grid is 2.11$\times$, 2.05$\times$, and 1.90$\times$ faster to build than the closest competitor, STRtree. Quadtree is the slowest to build because it generates a large number of cells for optimal configuration. Not all partitions in Quadtree contain an equal number of points as it divides the space rather than the data, thus leading to an imbalanced number of points per partition. Fixed-grid and adaptive-grid do not generate a big number of partitions, as the partitions are quite large for optimal configuration. They are smaller for similar reasons. The index size in Figure~\ref{fig:indexing_costs} also includes the size of the indexed data.

In the figure, we can also see that the learned indexes are faster to build and consume less memory than the S2PointIndex and JTS STRtree. Fixed-grid, for example, is from 2.34$\times$ to 15.36$\times$ faster to build than S2PointIndex, and from 11.09$\times$ to 19.74$\times$ faster to build than JTS STRtree. It also consumes less memory than S2PointIndex (from 3.04$\times$ to 3.4$\times$) and JTS STRtree (from 4.96$\times$ to 8.024$\times$).
However, we note that the comparison of the index size with JTS STRtree is not completely fair. 
JTS STRtree is a SAM (spatial access method), where it stores four coordinates for each point (since the points have been stored as degenerate rectangles). The learned indexes implemented in this work are PAMs (point access method), where we only store two coordinates for each data point.

\greybox{\textbf{Key Takeaways.} Grid-based indexes are faster to build and consume less space compared to tree-based indexes. 
Optimally-tuned grid-based indexes use larger partitions, which results in a small total number of partitions.
In contrast, tree-based indexes produce a large number of smaller partitions. 
Since we embed learned models within each partition, tree-based indexes have a higher build time and are larger in size, as they maintain more learned models.}
\section{Related Work}
\label{sec:relatedwork}
Recent work by Kraska et al.~\cite{rmi} proposed the idea of replacing traditional database indexes with learned models. 
Since then, there has been a corpus of work on extending the ideas of the learned index to spatial and multidimensional data.

\textbf{Learned Multidimensional Indexing and Partitioning.} Flood~\cite{nathan2020flood} is an in-memory read-optimized multidimensional index that organizes the physical layout of $d$-dimensional data by dividing each dimension into some number of partitions, which forms a grid over the $d$-dimensional space and adapts to the data and query workload. 

Similarly to Flood, our implementation of the grid indexes partitions the data using a grid across $d-1$ dimensions and uses the last dimension as the sort dimension.
Tsunami\cite{tsunami, tsunami_2} is an improvement over Flood, designed to efficiently handle correlated and skewed data. 
Machine learning techniques have also been applied to reduce the I/O cost for disk-based multidimensional indexes. 
Qd-tree~\cite{qdtree-sigmod} uses reinforcement learning to construct a partitioning strategy that minimizes the number of disk-based blocks accessed by a query. 
The ZM-index~\cite{zm-index} combines the standard Z-order space-filling curve with the Recursive-Model Indexes (RMI) proposed by Kraska et al.~\cite{rmi} by mapping multidimensional values into a single-dimensional space that is learnable using models. 
The ML-index~\cite{ml-index} combines the ideas of iDistance~\cite{idistance} and RMI~\cite{rmi} to support range and KNN queries. 
There are also efforts to augment existing indexes with lightweight models to accelerate range and point queries~\cite{hands_off}. 
Machine learning has also been applied to various other aspects of data management~\cite{learned_lsi, learned_scheduler_1, flirt, learned_instance_optimized_z_1, learned_hash, learned_joins}.

\textbf{Learned Spatial Indexes and Algorithms.} LISA~\cite{lisa-sigmod} is a disk-based learned spatial index that achieves low storage consumption and I/O cost while supporting range and nearest neighbor queries, insertions, and deletions. In~\cite{instance_optimal_z_curve}, the authors propose an instance-optimized Z-curve index. 
They present alternate ordering and partitioning of the Z-curve and propose two greedy heuristics that learn the most effective ordering for a particular workload and dataset.

In traditional R-trees, the branches visited for a particular range query depend on the overlap of the query with multiple children nodes. In~\cite{ai_r_tree}, the authors propose using the overlap ratio (the required number of leaf nodes / actually visited leaf nodes) to identify high-overlap queries. They also build an AI tree that uses uniquely assigned IDs to the R-tree leaf nodes as class labels for multi-label classification. At runtime, their technique identifies queries with high overlap and uses the AI tree for them; otherwise, it falls back to the regular R-tree for low-overlap queries. The AI tree helps minimize the number of visited leaf nodes.
In~\cite{glin}, the authors work on SAMs (i.e., spatial objects with extent such as linestrings and polygons). 
The authors compute the Z-curve extent (minimum and maximum of the Z-interval) of the MBR enclosing the spatial object, sort the geometries based on the Z-address interval and build a hierarchical tree-like structure where internal nodes contain the linear regression model and an array of pointers to child nodes, while the leaf node contains the linear regression model and an array of actual data along with the MBR of the data. This is in contrast to the ZM-index which uses Z-curve values of the points instead of spatial object with extents, as well as the Hibert-curve index that we use to store the points. 
SPRIG~\cite{sprig} is a spatial interpolation function-based grid index. The authors use spatial bilinear interpolation function to predict the position of the query points and then use a local binary search to refine the result. 
Spatial Join Machine Learning (SJML)~\cite{cost_model_spatial_join,  cost_model_spatial_join_2} is a machine learning-based query optimizer for distributed spatial joins. 
It consists of three levels: (1) the first level builds the cardinality estimates model that learns the result size of the spatial join, (2) the second level combines the predicted result size with other dataset characteristics to build a separate model which predicts the number of geometric comparison operations, and (3) builds a classification model which is able to predict the best join algorithm. 
In~\cite{rlr_tree}, the authors exploit the fact that the R tree can be constructed using multiple splitting strategies, e.g., linear split, quadratic split, R*-tree split etc. 
They run a microbenchmark to find that the query performance varies for various query workloads and datasets. They identify that ChooseSubtree and Split operations for the tree construction can be considered as sequential decision-making problems. They model them as two Markov Decision Processes (MDP) and use reinforcement learning to learn the model for the two operations.

Machine Learning techniques on spatial data have also been applied to various scenarios such as spatio-textual queries~\cite{learned_spatial_textual_1}, social media data~\cite{learned_social_media_1}, passage retrieval~\cite{learned_high_dim_1}, and streaming~\cite{learned_streaming}, and other areas~\cite{learned_lsi, learned_scheduler_1, flirt, learned_instance_optimized_z_1}. There also exists various surveys that cover in-depth various works that apply machine learning to spatial data~\cite{ml_spatial_survey_1, ml_spatial_survey_2, ml_spatial_survey_3}.
\section{Conclusions and Future Work}
\label{sec:conclusion}

In this work, we implemented learning-enhanced variants of six classical spatial indexes.
We found that, in most cases, the fixed-grid is the best-performing learning-enhanced index and also the simplest one to implement. 
Next, we summarize the results based on the four queries covered in this work and discuss avenues for future work.

\textbf{Range Queries.} Recent advancements in applying machine learning to databases have focused on creating instance-optimized~\cite{instance_optimized_1, instance_optimized_2} versions of various components, including index structures. Motivated by this, in Section~\ref{sec:range_query_tuning} we showed that the performance of various index structures fluctuates depending on the query workload and dataset. We demonstrated that tuning the index structures based on both the dataset and the query workloads is essential to obtain optimal performance. Additionally, we showed that using learned models instead of binary search improves performance by 11-39\%.

We also found that learned models do not help much in the case of tree-based index structures.
The performance gains were minimal, ranging from 2\% to 9\%, and in some instances, learned models even resulted in worse performance.
We showed that fixed-grid was the best-performing index for range queries, outperforming the closest tree-based competitor by a factor of 1.23$\times$ up to 1.83$\times$. This is in \emph{contrast} to traditional knowledge that grid-based structures suffer from skewed partitions. 
We argue that when all data can be indexed and stored in main memory, grid-based index structures, when tuned for optimal partition sizes, outperform tree-based index structures.
We also showed that fixed-grid is from 8.67$\times$ to 43.27$\times$ faster than JTS STRtree and from 4.34$\times$ to 53.34$\times$ faster than S2PointIndex. 
Additionally, we addressed the reasons behind the poor performance of Hilbert curve-based approaches, i.e., S2PointIndex, and learned linearized Hilbert curve-based index.

\textbf{Point Queries.} In Section~\ref{sec:point_query_performance}, we again observe that fixed-grid is the best-performing index. 
It is 1.51$\times$ to 2.27$\times$ faster compared to the closest tree-based competitor, kdtree. 
We also showed that the learned index based on linearized Hilbert curve values is highly competitive, as point queries require searching for a specific point on the Hilbert curve. 
This is in contrast to range queries where it may end up scanning a lot of redundant points.

\textbf{Distance Queries.} The fixed-grid index has again the best performance for distance queries. 
However, we note that, as in the case of high selectivity range queries, the performance gains diminish with increasing selectivity, and the Haversine distance computation becomes the \emph{dominant} cost. 
We also observe that S2PointIndex performs the best in the case of uniform queries in the OSM dataset. There are two reasons for this: (1) the filter phase produces a lot more candidate points in the case of uniform queries compared to other datasets and query workloads, and (2) S2PointIndex traverses the Hilbert curve in an efficient manner avoiding the overheads of the \emph{filter and refine} approach.

\textbf{Join Queries.} The \emph{filter} phase of the join, which internally issues a range query to the index structures using the bounding box of the polygons, exhibits the same performance as range queries. 
However, the \emph{refinement} phase of the join is the most \emph{dominant} cost. This leads to similar performance of the learned index structures and performance gains are limited. 
Currently, we utilize interval trees to index the polygons used in the join operation. 
In the future, we plan to use learned index structures, such as those proposed in~\cite{bounded_approx} and~\cite{glin}, to index the polygons as well.

\textbf{Future Work.} Thus far, we have only studied the case where both the indexes and data fit into RAM.
For disk-based use cases, the performance will likely be dominated by I/O, and the search within partitions will be less important.
We expect the partition sizes to be performance-optimal when aligned with the physical page size.
To reduce I/O, it will be crucial to eliminate any unnecessary points from the partitions.
Therefore, we anticipate that using two-dimensional indexing will be the best approach for disk-based storage.
For further discussion on this topic, we refer to LISA~\cite{lisa-sigmod}.
Our findings demonstrate that grid-based index structures outperform tree-based indexes by reducing random accesses and using efficient fast search over large partitions.
However, the gains diminish when computationally expensive operations, such as Haversine distance computation and point-in-polygon tests, are required. 
To minimize the overall query runtime, it is necessary to develop novel effective indexing methods for polygons (to minimize or eliminate point-in-polygon tests) and circles (to avoid Haversine distance computation).
Moreover, in our work, we tuned the index structures using a manual process. In the future, our aim is to investigate the use of deep learning-based methods to tune the index structures for various types of queries, datasets, and workloads.

\balance

\bibliographystyle{ACM-Reference-Format}
\bibliography{learnedspatial}

\end{document}